\documentclass[
11pt,
aps,
pra,
tightenlines,
longbibliography,
showpacs,
nofootinbib,
notitlepage,
superscriptaddress]{revtex4-2}

\usepackage[title]{appendix}
\usepackage{graphicx, multirow}
\usepackage{amsmath, amssymb, physics}
\usepackage{booktabs}
\newcommand{\mlc}[1]{\begin{tabular}[c]{@{}l@{}}#1\end{tabular}}

\usepackage{hyperref}
\usepackage[capitalize]{cleveref}
\usepackage{siunitx}

\usepackage{caption}
\usepackage{subcaption}
\begin{document}

\title{Impacts of Decoder Latency on a Utility-Scale \\ Quantum Computer Architecture}

\author{Abdullah~Khalid}
\author{Allyson~Silva}
\affiliation{1QB Information Technologies (1QBit), Vancouver, BC, Canada}
\author{Gebremedhin~A.~Dagnew}
\affiliation{1QB Information Technologies (1QBit), Vancouver, BC, Canada}
\affiliation{Department of Physics \& Astronomy, University of Waterloo, Waterloo, ON, Canada}
\affiliation{Perimeter Institute for Theoretical Physics, Waterloo, ON, Canada}

\author{Tom Dvir}
\author{Oded Wertheim}
\author{Motty Gruda}
\affiliation{Quantum Machines, Israel}

\author{Xiangzhou~Kong}
\author{Mia Kramer}
\author{Zak~Webb}
\author{Artur~Scherer}
\thanks{{\vskip-10pt}{\hskip-8pt}Corresponding authors: \href{mailto:pooya.ronagh@uwaterloo.ca}{pooya.ronagh@uwaterloo.ca}, \href{mailto:artur.scherer@1qbit.com}{artur.scherer@1qbit.com}\\}
\affiliation{1QB Information Technologies (1QBit), Vancouver, BC, Canada}

\author{Masoud~Mohseni}
\affiliation{HPE Quantum, Hewlett Packard Labs, CA, USA}
\affiliation{Emergent Machine Intelligence, Hewlett Packard Labs, CA, USA}

\author{Yonatan~Cohen}
\affiliation{Quantum Machines, Israel}

\author{Pooya~Ronagh}
\thanks{{\vskip-10pt}{\hskip-8pt}Corresponding authors: \href{mailto:pooya.ronagh@uwaterloo.ca}{pooya.ronagh@uwaterloo.ca}, \href{mailto:artur.scherer@1qbit.com}{artur.scherer@1qbit.com}\\}
\affiliation{1QB Information Technologies (1QBit), Vancouver, BC, Canada}
\affiliation{Department of Physics \& Astronomy, University of Waterloo, Waterloo, ON, Canada}
\affiliation{Perimeter Institute for Theoretical Physics, Waterloo, ON, Canada}
\affiliation{Institute for Quantum Computing, University of Waterloo, Waterloo, ON, Canada}
\affiliation{Microsoft Quantum, Microsoft Corporation, WA, USA}

\date{August 21, 2026}


\begin{abstract}
The speed of a fault-tolerant quantum computer depends in large part on the reaction time of its classical electronics, that is, the total time required by decoders and controllers to determine the outcome of a logical measurement and execute subsequent conditional logical operations. Despite its importance, the reaction time and its impact on the design of a logical microarchitecture of a quantum computer are not well understood. In this work, we perform a detailed system-level analysis of the classical and quantum resource requirements of a surface code based architecture. To this end, we construct a model of the reaction time in which the decoder latency is based on parallel space- and time-window decoding methods. In addition, we draw communication latencies from our envisioned quantum execution environment, which comprises a high-speed network of quantum processing units, controllers, decoders, and high-performance computing nodes. We use this model to estimate the increase in the logical error rate of magic state injections as a function of the reaction time. We then show how the microarchitecture can be optimized with respect to the reaction time, and present full-system quantum resource estimates based on realistic hardware noise parameters for executing utility-scale quantum circuits of the \mbox{Fermi--Hubbard} model (2562 logical qubits and $4$$\times$$10^6$ $T$ gates) and NMR spectral prediction (241 logical qubits
and $5.11$$\times$$10^{11}$ $T$ gates). We numerically determine the impact of improving reaction time on the code distance required for the core processor and on the size of the resource state factory. Moreover, we analyze how decoder speeds constrain the size of practically executable circuits, and show that state-of-the-art decoders will require a performance improvement of at least an order of magnitude to execute utility-scale quantum algorithms within days.
\end{abstract}

\maketitle

\section{Introduction}

Fault-tolerant quantum computation (FTQC) is an inherently hybrid quantum--classical scheme that involves executing quantum circuits, measuring subsets of the qubits involved in the circuits to determine error syndromes, classically inferring the most-probable errors from the syndromes, and repeating this process iteratively. Utility-scale quantum programs require performing trillions of operations with thousands of logical qubits~\cite{Mohseni2026How}. Therefore, their corresponding fault-tolerantly compiled physical circuits require millions of qubits to operate robustly in communication with classical control and decoding electronics for many hours, days, or longer. Building such a complex machine presents many systems engineering challenges. In this paper, we focus on the challenges caused by the delays produced by this hybrid process and discuss how system and software architectures can be adapted to minimize the impact of these delays on the performance of a quantum computer.
 
The most significant delay in a quantum system occurs when conditional logical operations in a circuit have to wait for the outcome of preceding logical measurements to be determined. This wait, called the \emph{reaction time}, results from the sum of the decoder and the control electronics latencies in the system. The reaction time determines the speed of a quantum computer~\cite{Gidney2019Flexible}; hence, minimizing the reaction time is critical for fast qubits (superconducting, topological, spin, etc.); otherwise, the low gate and measurement times of these qubits are negated by slow decoders. Consequently, substantial effort has been made towards developing faster decoders, including implementations on FPGAs and ASICs~\cite{Delfosse2021Almostlinear, Das2022AFS,  Higgott2021PyMatching, Higgott2025Sparse, Battistel2023Realtime, iOlius2024Decoding, Bausch2024Learning, Barber2025Realtime, Muller2025Improved, Maurya2025FPGAtailored}. However, none of these are fast enough to decode spatially or temporally large logical operations while matching syndrome throughput, a key requirement for FTQC~\cite{Terhal2015Quantum}. To solve this issue, several decoder parallelization algorithms have been developed that allow significant improvements in both decoding throughput and latency~\cite{Bombin2023Modular, Skoric2023Parallel, Tan2023Scalable, Chan2024Snowflake, FuhuiLin2025Spatially, Zhang2025LATTE, Zhang2026Learning}.

Most studies on quantum resource estimation have ignored the mentioned complications with decoders, and instead assumed a fixed and small reaction time~\cite{Gidney2021How, Lee2021Even, Litinski2022Active, Gidney2025How}. Counter to this, in a recent study~\cite{Silva2025Optimizing}, some authors of this work discuss how the \emph{logical microarchitecture} of surface code based quantum processing units (QPU) can be optimized with respect to reaction time. A logical microarchitecture is an abstract partitioning of a QPU into dedicated areas for various fault-tolerant procedures, and is crucial for creating the appropriate abstraction layers within the compilation stack of the computer.

In this work, we extend our analysis of the impact of reaction time on the software and the system architecture of a quantum computer in several ways. First, we use a method that combines parallel space- and time-window decoding techniques~\cite{Skoric2023Parallel,FuhuiLin2025Spatially}, to show that out of the two reaction times in the microarchitecture presented in Ref.~\cite{Silva2025Optimizing} only one controls the speed of a computation. Second, we build a model of the additional logical error accumulated due to the delays induced by the two reaction times, and use this model to estimate the required number of physical qubits as a function of the reaction times for two utility-scale quantum algorithms. Third, we develop decoder latency models for decoding quantum memories and lattice surgeries~\cite{Fowler2012Surface, Horsman2012Surface}. We also experimentally estimate the communication latencies in a \emph{quantum execution environment} we have designed. The execution environment is a network between a controller, a high-performance decoder cluster, and an \emph{orchestrator} that triggers the execution of quantum applications (see \cref{fig:system-architecture}). Moreover, we identify target performance metrics for decoders that enable the execution of quantum algorithms within a threshold ranging from one hour to one month. Finally, we estimate the number of decoding units required for a QPU with 10M qubits.

This paper is organized as follows. In \cref{sec:background}, we review background work on surface codes and their decoding, fault-tolerant quantum compilation and assembly, and the optimization of logical microarchitectures. In \cref{sec:coprocessing}, we describe our envisioned quantum execution environment for a quantum computer, highlighting the flow of instructions and information between the classical and quantum processing units. We also present experimentally determined numbers for the communication latency between different units. In \cref{sec:reaction_time}, we analyze the reaction time for a magic state injection protocol and the resulting decoding jobs required to be completed at runtime. We provide the results of our system-level analysis in \cref{sec:results-resource-estimates,sec:decoder-requirements}. In \cref{sec:results-resource-estimates}, we show the impact of reaction time on optimal microarchitectures and physical qubit counts. We identify decoder speed requirements and the number of decoders necessary for utility-scale quantum computing in \cref{sec:decoder-requirements}. Finally, we provide our concluding remarks in \cref{sec:conclusion}.

\section{Background}
\label{sec:background}

\subsection{Fault-Tolerant Quantum Computation Using Surface Codes}
\label{sec:surface-codes}

In this paper, we study fault-tolerant quantum architectures based on the rotated surface code~\cite{Dennis2002Topological,Fowler2012Surface,Horsman2012Surface}. In the simplest case, a $d \times d$ patch of data qubits can be used to encode a single logical qubit with an error-correcting distance of $d$. However, others shapes and configurations are possible; for example, one can also encode two logical qubits in a $2d \times d$ patch of data qubits such that each logical qubit has a distance of $d$~\cite{Fowler2012Surface,Litinski2018Lattice,Litinski2019Game}. Segments of the geometric boundaries of these patches define the logical Pauli operators of the logical qubits.

A variety of operations may be performed on the surface code patches that are laid out in a 2D geometry~\cite{Fowler2019Low}. Single-qubit Pauli operations and their measurements can be performed by acting directly on the data qubits. Single-qubit patches may also be grown into larger patches using a growth protocol. Multi-qubit operations, however, are performed using \emph{lattice surgery}, in which interactions between logical qubits on distinct code patches are engineered by first merging them into a single patch and then splitting them back into their original patches~\cite{Horsman2012Surface, Litinski2018Lattice, Erhard2021Entangling, Chamberland2022Universal}. This process leads to a joint measurement of the encoded logical qubits in the joint basis of the multi-qubit Pauli operator $Q$ associated with the merged boundaries of the patches. For example, if the $X$ boundaries of two patches in the logical $\ket{00}_L$ state are merged and split, the qubits are prepared in the Bell state
\begin{equation}
\frac{1}{\sqrt{2}}\Big[I + mX_{L,0}X_{L,1}\Big]\ket{00}_L = \frac{1}{\sqrt{2}}\Big[\ket{00}_L + m\ket{11}_L\Big], \label{eq:lattice_surgery_outcome}
\end{equation}
where $m \in \{+1,-1\}$ is the outcome of the measurement, determined by multiplying the values of all $X$ syndromes in the bus region connecting the two patches. This procedure constitutes what is referred to as an $X \otimes X$ surgery. Most logical operations are executed in $d$ stabilization rounds, which we call the \emph{logical cycle time}, $\tau_\text{logical}$.

\begin{figure}
    \centering
    \includegraphics{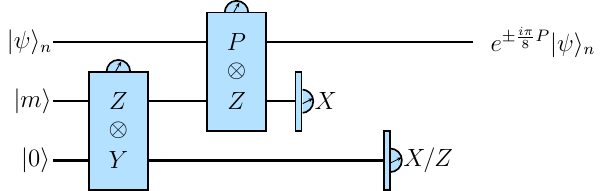}
    \caption{Circuit for a post-corrected $\pi/8$ rotation gadget~\cite{Litinski2019Game}. An $n$-qubit state $\ket{\psi}_n$ can be rotated by an angle of $\pm\pi/8$ around an $n$-qubit Pauli $P$ with the help of a magic state, an ancilla called \emph{correction qubit}, two non-destructive multi-qubit measurements (rectangular boxes), and two destructive single-qubit measurements (thin rectangles). The correction qubit can be measured in either the $X$ or $Z$ basis, and the sign of the rotation angle is determined by this choice of basis and the value of the $P \otimes Z$ measurement. A Pauli correction may be required depending on the value of the measurement outcomes. See Appendix~\ref{sec:rotation-gadget} for further details. If this gadget is executed on logical qubits encoded in the surface code, the multi-qubit measurements are performed in one logical cycle (which we take to be $d$ stabilization rounds) each using lattice surgery, and the single-qubit measurements are performed in zero time by combining them with the preceding logical operation (i.e., lattice surgery or memory).}
    \label{fig:pi8_gadget}
\end{figure}

Multi-qubit Pauli measurements, along with ancilla prepared in specific states, can be used to engineer nontrivial multi-qubit operations, such as a rotation by an angle $\theta = \pi/4$ or $\pi/8$ around the axis of Pauli $P$, that is, the interaction $\ket{\psi} \to e^{i\theta P}\ket{\psi}$~\cite{Litinski2019Game}. A quantum algorithm can be compiled into a sequence consisting solely of multi-qubit $\pi/8$ rotations. Such a a method of computation is an example of Pauli-based computation~\cite{Bravyi2016Trading,Chamberland2022Universal}. A circuit from Ref.~\cite{Litinski2019Game} implementing a $\pi/8$ rotation, called the post-corrected \emph{$\pi/8$ rotation gadget}, is depicted in \cref{fig:pi8_gadget}. The first step in this circuit involves a magic state $\ket{m}$ being entangled with a \textit{correction qubit} initialized in $\ket{0}$ using a $Z\otimes Y$ lattice surgery. Then, a $P\otimes Z$ surgery is used to entangle the magic state qubit with the $n$ computation qubits initially in the state $\ket{\psi}_n$. Next, the magic state qubit is measured in the $X$ basis. At this point, the state may be rotated by either $\pi/8$ or $-\pi/8$ depending on the random outcome of the $P \otimes Z$ surgery and the basis in which the correction qubit is measured. The choice of the rotation angle basis ($X$ or $Z$) depends on the measurement outcomes of correction qubits in preceding $\pi/8$ rotations that anti-commute with $P$. Additional information is provided in Appendix~\ref{sec:rotation-gadget}.

\subsection{The Fault-Tolerant Logical Microarchitecture}
\label{sec:ftqc-assembly}

\begin{figure}[tb]
    \centering
    \includegraphics[width=0.95\textwidth]{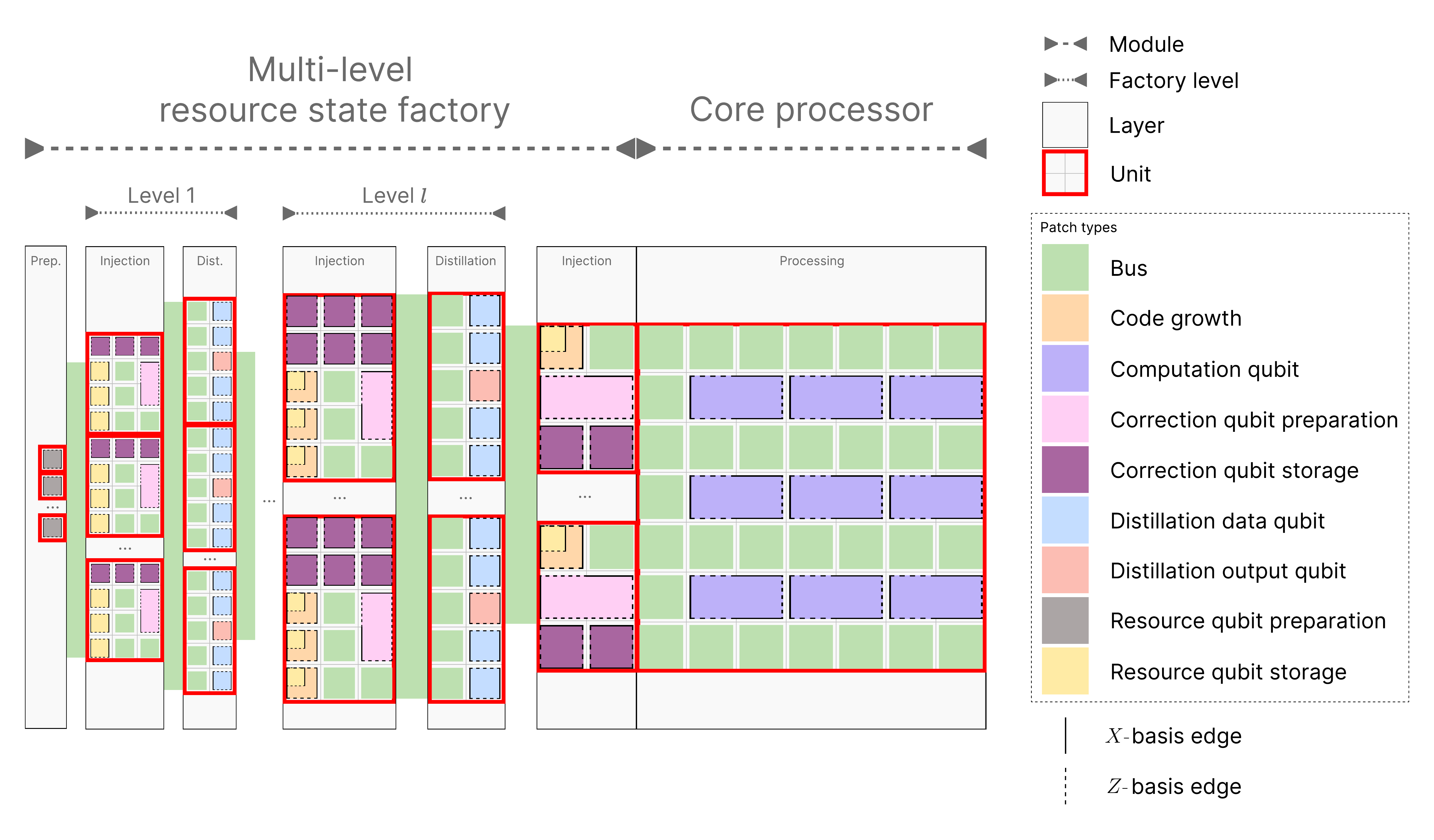}
    \caption{An instance from the \emph{Litinski-style monolithic planar surface code} (LMPSC) logical microarchitecture family~\cite{Silva2025Optimizing}. Physical qubits are divided into fixed rectangular code patches of varying distances representing logical qubits. Patches form functional units; units with the same function form layers; layers nest within modules. Bus regions facilitate lattice surgery between distant logical qubits. The microarchitecture depicted has two modules: a \emph{core processor}, where the algorithm is executed on computation qubit patches that store pairs of logical qubits, and a \emph{resource state factory} (RSF), where high-fidelity magic states are produced across layers grouped into factory levels. In the RSF, a \emph{preparation} layer prepares magic states in single patches using a preparation protocol. If fidelity is insufficient, magic states pass through distillation rounds that progressively improve magic state fidelities. Each round comprises a factory level with: an \emph{injection} layer, comprising units implementing the circuit depicted in \cref{fig:pi8_gadget} while providing additional space for growing code distances of incoming magic states and storing the correction qubits waiting for the measurement outcomes from the decoder; and a \emph{distillation} layer, here implementing the 15-to-1 distillation protocol~\cite{Litinski2019Game}. A final injection layer interfaces with the core processor. Unit layouts between same-purpose layers may vary with required throughput. This instance shows examples of injection units ranging from compact-but-slower to larger footprints enabling more parallelization via an extended bus. Code distances also vary across the microarchitecture, as reflected in patch sizes, leading to differing logical cycle times across layers.}
    \label{fig:architecture_with_correction_storage}
\end{figure}

The logical microarchitecture of a quantum computer is the specific arrangement of logical modules and submodules in the QPU that enable the execution of a quantum program. Many different microarchitectures may be designed to execute the same program, possibly requiring different space and time resources. Litinski demonstrates how a quantum algorithm can be compiled to rotation gadgets realized via Pauli product measurements, and introduces a surface code based microarchitecture family to execute the compiled program~\cite{Litinski2019Game}. In Ref.~\cite{Silva2025Optimizing}, this family of microarchitectures is significantly restructured, such as by overhauling the design of the magic state production area, and introducing dedicated regions for injection and storage of magic states. Here, we call this family the \emph{Litinski-style monolithic planar surface code} (LMPSC) logical microarchitecture family, which is suitable for the execution of algorithms compiled to post-corrected $\pi/8$ rotation gadgets. An instance of the LMPSC microarchitecture is depicted in \cref{fig:architecture_with_correction_storage}. We describe each microarchitecture in this family as a functional hierarchy of modules, levels, layers, units, and patches. At the highest level, each microarchitecture is composed of two modules: a \emph{core processor} where the logical computation is executed and a \emph{resource state factory} (RSF) that periodically produces high-quality magic states. These modules themselves comprise layers with distinct functions, with the RSF also comprising levels implementing layers for rounds of distillation to improve magic state fidelity. Different code distances may be used across modules and levels. Each layer consists of one or more copies of the same unit, each of which performs a specific task, examples of which are given below. A unit consists of a set of equal-distance logical surface code patches, each of which has a specific purpose. Throughout the microarchitecture, bus regions are used to facilitate lattice surgery between distant qubits, which includes moving magic states between layers until reaching the core processor.

The core processor module consists of a \emph{processing} layer with a single unit, assuming the computation is performed on a single QPU. This unit stores the logical qubits associated with the algorithm. These qubits are stored in pairs in $2d \times d$ sized patches called \emph{computation qubit} patches. This layout, called \textit{fast block} by Litinski~\cite{Litinski2019Game}, minimizes the required bus area while leaving the $X$- and $Z$-Pauli operators of both logical qubits accessible. The RSF is composed of a \emph{preparation layer} and, optionally, \emph{factory levels} to implement distillation rounds. The preparation layer consists of units with a single patch, called \emph{resource qubit preparation}. Each unit prepares magic states using a selected magic state preparation protocol~\cite{Gidney2023Cleaner,Gidney2024Magic, Claes2025Cultivating, Chen2025Efficient, Sahay2026Foldtransversal}, possibly employing space or time parallelization~\cite{Chung2026Partially}. Successfully created magic states are moved to a \emph{resource qubit storage} patch in a unit in the next injection layer, which either directly interfaces with the core processor or, if the magic state fidelity is too low, is within the first factory level. In an injection unit, the magic state distance is grown using \emph{code growth} patches, if required. The injection unit is used to implement the post-corrected $\pi/8$ rotation gadget depicted in \cref{fig:pi8_gadget}. The correction qubit is prepared in a \emph{correction qubit preparation} patch used to perform the rotation gadget's $Z \otimes Y$ lattice surgery with the magic state. \emph{Correction qubit storage} patches are allocated to store correction qubits waiting for the measurement outcomes from the decoder. When in a factory level, the low-fidelity magic state is injected into a unit in the distillation layer. Each distillation unit implements the \mbox{15-to-1} distillation protocol~\cite{Litinski2019Game}, which performs 11 $\pi/8$ rotations on five logical qubits. Four of these are stored in \emph{distillation data qubit} patches, while the \emph{distillation output qubit} patch holds the created magic state qubit. The injection layer that interfaces with the processing layer may be composed of one or more units based on the parallelization of the $\pi/8$ rotations in the compiled circuit. Different layouts may be used for units within a layer based on the desired space and time efficiency for distilling and injecting states.

This modular configuration of the microarchitecture with the preparation, distillation, and injection protocols performed in dedicated units allows the magic state and correction qubit resource flows to be resized by simply defining the number and size of the structures required by them. The necessary number of distillation levels is dictated by the target magic state error rate relative to what a single distillation round can achieve. For the \mbox{15-to-1} protocol, the output error rate of one round scales as $e_\text{out} = \mathcal{O}(35\, e_\text{in}^3)$ in the input magic state error rate $e_\text{in}$~\cite{Litinski2019Game}. Taking $e_\text{in}\approx 4.73\times10^{-5}$ to be representative of the magic state cultivation's error rate on near-term transmon hardware (see Appendix~\ref{sec:ftqc-performance}), a single round yields $e_\text{out} \approx 3.7\times10^{-12}$. Any target magic state error budget below this value cannot be met with one distillation level, irrespective of unit layout or parallelization. For a typical overall error budget of $1\%$, this threshold corresponds to circuits with $T$ counts on the order of $10^9$ or more, before accounting for decomposition and memory error contributions. Consequently, deeper distillation hierarchies are a structural consequence of the target error budget and compiled $T$ count rather than an optional design choice. As $T$ counts in quantum algorithm benchmarking continue to decrease and magic state preparation error rates improve, single-level factories are becoming increasingly common in the microarchitecture literature~\cite{Gidney2025How, Chatterjee2025QSpellbook, Lee2025LowOverhead, Webster2026Pinnacle}. The number of distillation units at each level is determined by matching the magic state supply rate of that level to the demand rate of the next level or the core processor. In a distillation cycle of 12 higher-level logical cycles, each unit requires 15 input magic states for every output magic state it produces. However, the ratio of units between adjacent levels is typically smaller than 15 because lower-level units generally operate at smaller code distances, resulting in shorter logical cycle times and therefore higher magic state production rates.

The assembly of the quantum circuit onto this microarchitecture is achieved by first allocating a total logical error budget across the components that contribute logical errors along the execution path. The error contributions associated with the operations performed in the microarchitecture we consider are the following: (i) memory errors on idling computation and storage qubits; (ii) lattice surgery errors during multi-qubit measurements in the core processor; and (iii) RSF-induced errors, comprising preparation infidelity, logical faults within distillation rounds, and inter-level code growth. These contributions accumulate approximately additively over the sequence of logical events in a compiled schedule. Thus, meeting the error budget requires choosing code distances and capacities (layout and number of units in each layer) such that the sum of logical error contributions remains within the error budget.

Within this feasible region, optimal microarchitectures are those that minimize the space and/or time overhead. The time-optimal design is achieved when the RSF produces magic states at the same rate that the core can consume them. Faster decoders reduce the time spent waiting on instructions for the correction qubit, thereby allowing higher consumption rates. Even with decoder responses faster than logical cycle times, reaction time is the limiting factor in the post-corrected model since non-commuting $T$ gates can be parallelized via quantum teleportation~\cite{Fowler2013Timeoptimal}. Space-optimal designs deliberately create undersized RSFs to trade runtime for a smaller physical footprint. However, if undersized too aggressively, idling grows, memory error increases, and higher code distances may be required to stay within the budget, which may lead to a higher space overhead instead~\cite{Silva2025Optimizing}. Decoder latency directly affects this balance. Each correction qubit produced by the $\pi/8$ gadget must remain stored until the classical controls return the instructions for the measurement basis for that qubit. As soon as a correction qubit is measured, its storage patch is released and can be reused.

For distillation protocols where all $\pi/8$ rotations are commutable (e.g., the \mbox{15-to-1} protocol), the decision on the basis of each correction qubit measurement depends only on measurement outcomes within its own $\pi/8$ gadget. There are no cross-gadget dependencies. Consequently, when a correction qubit is prepared, it must be held in storage until the decoder processes the last dependent measurement for that gadget. In the steady state, this creates a pipeline in which a correction qubit enters storage roughly every logical cycle and remains stored for $\lceil{\gamma_\text{QM} / \tau_\text{logical}\rceil}$ logical cycles, which represents the reaction time $\gamma_\text{QM}$ expressed as a multiple of logical cycle duration $\tau_\text{logical}$ (see \cref{sec:reaction_time} for the definition of $\gamma_\text{QM}$). With sufficient decoder parallelism, the peak number of concurrently idling qubits per distillation unit is therefore exactly this ratio. Thus, each unit must provision at least this number of correction storage patches to avoid stalls due to insufficient storage.

\subsection{Decoding the Surface Code}
\label{sec:parallel-decoding}

Due to its attractive error-correcting properties and the relative simplicity of its implementation, many decoders have been developed for the surface code (see Refs.~\cite{Battistel2023Realtime, iOlius2024Decoding} for recent reviews). Such decoders can be judged based on three key metrics: accuracy, throughput and latency. The \emph{accuracy} of a decoder characterizes its ability to identify errors. The \emph{throughput} is the rate at which a decoder processes syndromes. If this is less than the rate at which syndromes are produced, a quantum computation will slow down exponentially~\cite{Terhal2015Quantum}. The \emph{latency} is the duration between when syndromes are made available to a decoder and when it outputs a result.

Whereas surface code decoders have low logical error rates, they struggle to match the throughput of the syndrome data produced, especially on superconducting qubit-based QPUs where the stabilization round time is $\SI{1}{\micro\second}$. The state-of-the-art decoder for the surface code is an FPGA implementation of the Union-Find decoder by Riverlane, which has a decoding time per stabilization round of $\SI{1.0}{\micro\second}$ for a quantum memory with $d=30$~\cite{Delfosse2021Almostlinear, Das2022AFS,Barber2025Realtime}. Such a decoder is fast enough to decode a quantum memory with $d\le 30$ if the stabilization time on the QPU is equal to or greater than $\SI{1.0}{\micro\second}$. However, lattice surgeries used in our FTQC architecture can involve hundreds or even thousands of patches, which require correspondingly larger throughputs from the decoder. Large lattice surgeries, if processed by individual decoding units, also lead to such long latencies so as to render utility-scale quantum computing practically infeasible.

Fortunately, various methods have been developed recently to break up a single decoding task into several smaller tasks that can then be solved in parallel using a classical coprocessor~\cite{Bombin2023Modular, Skoric2023Parallel, Tan2023Scalable, Chan2024Snowflake, FuhuiLin2025Spatially, Zhang2025LATTE,Zhang2026Learning}. These methods can significantly reduce the throughput requirements of a single decoder unit, as well as the total latency of the decoding task, at the cost of more decoding units. In our work, we use two such parallelization algorithms to build decoder latency models~\cite{Skoric2023Parallel, FuhuiLin2025Spatially}. These algorithms break the decoding graph~\cite{Higgott2021PyMatching, Chamberland2022Universal, Gidney2021Stim, Wu2022Interpretation, Derks2024Designing} into overlapping regions called \emph{windows}, which are then decoded separately and their results combined.

\begin{figure}[tb]
    \centering
    \includegraphics{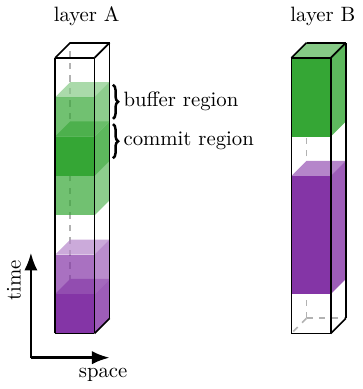}
    \caption{Decoding windows for the decoder parallelization algorithm for quantum memory. In this algorithm~\cite{Skoric2023Parallel}, the decoding graph is broken temporally into windows, which are shown here as coloured cuboids. Windows are assigned to one of two layers, and have commit regions (opaque in colour) and buffer regions (partially transparent in colour).}
    \label{fig:decoder_parallelization_memory}
\end{figure}

The decoding problem of a single surface code patch of distance $d$ stored in memory for many logical cycles can be solved by employing a strategy based on parallel windows in the time dimension of the decoding graph~\cite{Skoric2023Parallel}, illustrated in \cref{fig:decoder_parallelization_memory}. Sub-decoding tasks decode windows that have a spatial size of $d^2$ and a temporal size of $3d$, that is, $(x,z,t) = (d, d, 3d)$, and belong to one of two layers A and B. In layer A, the window for each task has three sub-regions, two buffer regions and a commit region, each of temporal size $d$. The windows associated with different tasks are separated by a gap of $d$ stabilization rounds. Once a layer-A decoding task is complete, the corrections made in the commit region are preserved, while the error strings extending from the commit region to the buffer regions are used to create artificial defects at the boundaries of the buffer regions. In layer B, the window for each task is a commit region of temporal size $3d$, consisting of the two buffer regions and the gap region between two adjacent windows from layer A. Hence, a layer-B task can be started only once the two adjacent layer-A tasks are completed. All tasks in a single layer can, in principle, be decoded in parallel. Hence, in the absence of parallelization overhead, if a single window is decoded in time $\tau_w$, then the decoder latency is $2\tau_w$.

\begin{figure*}[tb]
    \centering
    \begin{subfigure}[b]{0.49\textwidth}
        \centering
        \includegraphics{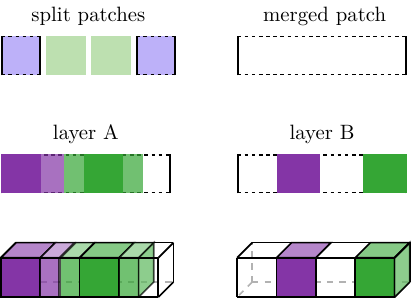}
        \caption{An $X\otimes X$ surgery}
        \label{fig:decoder_parallelization_ls_xz}
    \end{subfigure}
    \hfill
    \begin{subfigure}[b]{0.49\textwidth}
        \centering
        \includegraphics{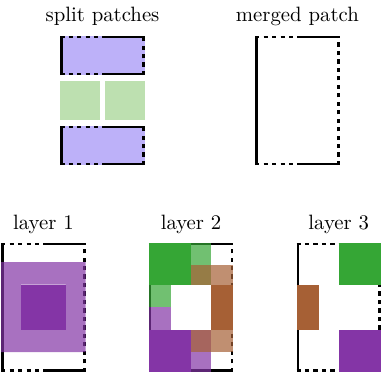}
        \caption{A $Y \otimes Y$ surgery}
        \label{fig:decoder_parallelization_ls_y}
    \end{subfigure}
    \caption{Decoding windows for the decoder parallelization algorithm for lattice surgery. In this algorithm~\cite{FuhuiLin2025Spatially}, the decoding graph of a single lattice surgery is broken spatially into windows, which are shown here as coloured rectangles (in the 2D cases) and cuboids (in the 3D cases). Windows are assigned to one of two or three layers, and have commit regions (opaque in colour) and buffer regions (partially transparent in colour). (a) Example of an $X \otimes X$ lattice surgery between two logical qubits connected by two bus patches. Depicted are the individual patches pre- and post-surgery (top left), the merged patch during the surgery (top right), the spatial slices of the windows in the two layers (middle), and the same windows shown in 3D (bottom). (b) Example of a $Y \otimes Y$ lattice surgery between two logical qubits (each encoded with another qubit in a rectangular patch) connected by a two-patch bus. Depicted are the individual patches pre- and post-surgery (top left), the merged patch during surgery (top right), and the spatial slices of the windows in the three layers (bottom).}
    \label{fig:decoder_parallelization_ls}
\end{figure*}

Decoding a lattice surgery between many patches can be accomplished using parallel windows in the space dimensions of the decoding graph~\cite{FuhuiLin2025Spatially}, illustrated in \cref{fig:decoder_parallelization_ls}. In this case, there are two layers, if the lattice surgery Pauli operator consists of only $X$ and $Z$ terms, and there are three layers if the Pauli operator has at least one $Y$ term. The windows in this method also contain commit and buffer regions. To ensure high accuracy, the commit regions of two windows in a single layer must not overlap, and the commit regions must be of spatial size $d \times d$. In the method presented in Ref.~\cite{FuhuiLin2025Spatially}, all windows have temporal size $d$ (the logical cycle time) because subsequent logical operations are decoded in sequence.

In the $Y$-surgery case, the first layer's windows have $1.5d$-sized buffers on all sides of the commit region, which results in windows of spatial size $2d \times 2d$. The windows in the second layer have spatial size $2d \times 1.5d$ because a portion of the decoding graph has already been committed. The third layer's windows consist of only commit regions of spatial size $d \times d$. To summarize, the three layers have windows of sizes $(2d, 2d, d)$, $(2d, 1.5d, d)$, and $(d, d, d)$, respectively.

\section{The Quantum Execution Environment}
\label{sec:coprocessing}

\begin{figure*}[tb]
    \centering
    \includegraphics[width=0.95\linewidth]{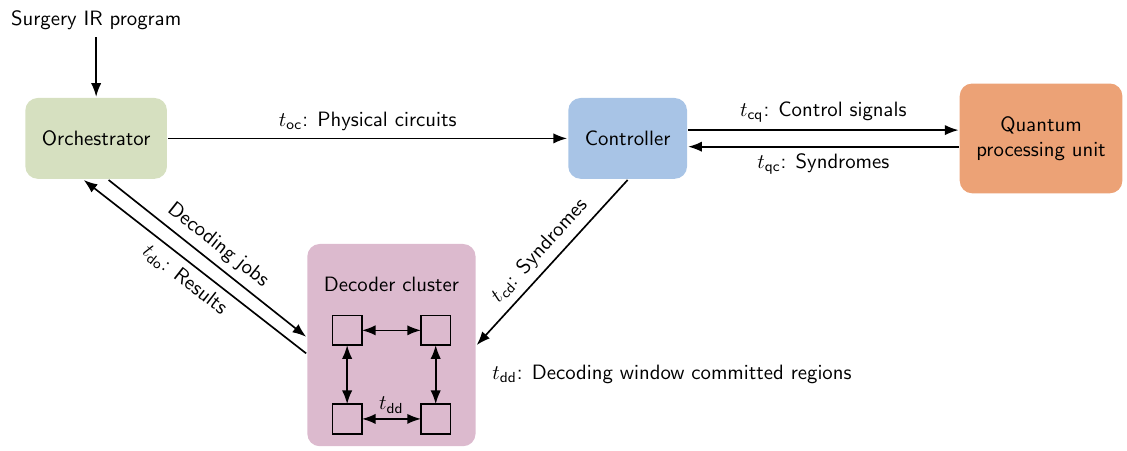}
    \caption{Schematic of the functional decomposition of the quantum execution environment. Arrows indicate the data transfer directions, and are labelled by the type of data and/or the corresponding communication times.}
    \label{fig:system-architecture}
\end{figure*}

A QPU's operations are executed with the help of a collection of classical processing units that form its quantum execution environment. The overall performance of a quantum computer depends on the speed of each component as well as the communication latencies between them. As explained earlier, the microarchitecture is also optimized with respect to these communication latencies. Here, we discuss our proposed execution environment, which is shown in \cref{fig:system-architecture}. We describe the classical processing units and estimate the speed of the parallel communication links in the environment. The three units comprise an orchestrator, a modular controller, and a decoder cluster.

The orchestrator is a classical computing device that serves as the execution provider. It receives a quantum program that has been compiled into surgeries on a collection of surface code patches, as described in \cref{sec:ftqc-assembly}~\cite{Litinski2019Game, Silva2024Multiqubit}. We call this the \emph{surgery intermediate representation} language, or surgery IR for short. The orchestrator parses the quantum program into a \emph{physical IR}, a representation of the program in the form of physical stabilizer circuits. The physical IR is sent to the controller with a latency of $t_\text{oc}$. The physical IR contains a large amount of data; for example, for a QPU with 10M qubits, the bandwidths of the links is in the Tb/s regime, which is well within capacity of present-day networking technology. The controller converts the physical IR into analog instructions that are sent to the QPU with a latency of $t_\text{cq}$. After the execution of a stabilization round, the syndrome qubits are measured and the results are transmitted from the QPU to the controller with a latency of $t_\text{qc}$, and from the controller to the decoder cluster with a latency of $t_\text{cd}$. Due to throughput improvements stemming from decoding parallelization, the total RAM required for storing syndrome measurement outcomes is on the order of gigabytes, as discussed in \cref{sec:decoder-how-many}.

The orchestrator also determines the decoding windows for each surgery and the sequence of all decoding jobs required for decoding each surgery along with their dependencies as explained in \cref{sec:parallel-decoding}. This information is communicated directly to the decoder cluster ahead of time, and thus does not contribute to the reaction time. We assume that the decoder cluster is equipped with an efficient workload manager that queues incoming decoding jobs and communicates dependencies between them. The dependencies are the boundaries of committed decoding regions from prior decoding windows and therefore comprise a small number of bits ($\sim$90 bits for distance 30 as calculated in Ref.~\cite{FuhuiLin2025Spatially}), communicated in a time of $t_\text{dd}$.

When the decoding of a surgery has completed, the results are transmitted to the orchestrator with latency $t_\text{do}$, which stores the physical and logical Pauli frames. The decoding results are used to update these frames and determine the results of logical measurements. The subsequent logical operations that are conditional on these logical measurements are updated by the orchestrator before they are parsed into the physical IR.

\subsection{Measurements of Data Transfer Time}
\label{sec:communication-latency}

We now present realistic estimates of the communication latencies expected in a future fault-tolerant quantum computer, which are later used to determine its total reaction time. For each latency depicted in \cref{fig:system-architecture}, we perform tests on state-of-the-art hardware and extrapolate the results to the scales necessary for utility-scale FTQC. These numbers depend critically on the volume of data transferred, the bandwidth of the corresponding communication channels, and the degree of parallelism (i.e., the number of concurrent channels) employed.

The principal data transfers are the transmission of the pulse sequences from the controller to the QPU and the return of the syndrome data from the QPU to the controller, both occurring over multiple analog channels and limited primarily by the signal time-of-flight. In addition, the transfer of syndromes to the decoder cluster requires dedicated digital communication links between the controller and the decoders.

In the modular control stack, each control cluster is assumed to manage on the order of $10^4$ physical qubits and to maintain independent communication links with both the decoder cluster and the orchestrator. This configuration determines the amount of data transmitted from the controller to the decoders in each stabilization round. Consequently, a 10M-qubit computer would consist of roughly $10^3$ such controller--decoder--orchestrator communication channels operating in parallel. 

We measure the round-trip latencies between an OPX1000 controller and an NVIDIA Grace Hopper server using an OPNIC PCIe optical interface, as a function of the data volume transmitted in each direction. We utilize the controller's state discrimination capabilities to reduce the required bandwidth to 1--2 bits per qubit. We observe a total communication time of under $\SI{2}{\micro\second}$ for sending and receiving 1000 bits of data (equivalent to a single logical patch) or 3--4 \SI{}{\micro\second} for a 10-patch surgery. Note that, as the system scales, the interface will be optimized accordingly with higher channel counts and data rates. Sending the data only to the relevant decoder in the cluster will minimize the impact on the latency resulting from large data transmissions. \cref{tab:data-transfer-time} presents the estimated time required for each data transfer. The values reflect the approximate duration per channel, representative data size, and operational periodicity.

\begin{table}[tb]
\centering
\begin{tabular}{ l  c  c  c }
\toprule
{Communication Step} & {Time ($\SI{}{\micro\second}$)} & {Size per Channel} & {Number of Channels} \\
\midrule
\mlc{$t_{\mathrm{qc}}$ -- Syndrome transfer \\\,\,\,\, from QPU to controller} & 0.15 & 1 & $5\times10^{6}$ \\
\mlc{$t_{\mathrm{cd}}$ -- Syndrome transfer \\\,\,\,\,from controller to decoders} & 2 & 5000 & 1000 \\
\mlc{$t_{\mathrm{dd}}$ -- Decoder-to-decoder\\ \,\,\,\,exchange} & 0.5 & 100 & multiple \\
\mlc{$t_{\mathrm{do}}$ -- Decoding results transfer\\\,\,\,\,from decoders to orchestrator} & 1 & 50,000 & 100 \\
\mlc{$t_{\mathrm{oc}}$ -- Instructions from\\\,\,\,\,orchestrator to controller} & 4 & 20,000 & 1000 \\
\mlc{$t_{\mathrm{cq}}$ -- Instructions from\\\,\,\,\,controller to QPU} & 0.15 & 1 & $5\times10^{6}$ \\
\bottomrule
\end{tabular}
\caption{For the quantum execution environment, the data transfer steps and their associated latency, the data size per channel, and the number of channels.}
\label{tab:data-transfer-time}
\end{table}

Note that the orchestrator--controller, controller--QPU, and controller--decoder data transfers occur every stabilization round, the decoder--decoder data transfers are executed once or twice every logical cycle, and the decoder--orchestrator data transfer occur once every logical cycle. Using the communication latency values in \cref{tab:data-transfer-time} and assuming a multi-channel architecture, the end-to-end communication time from the last measurement to the arrival of the updated instructions at the QPU is 
\begin{equation}
\tau_{\mathrm{com}} \approx
t_{\mathrm{qc}} + t_{\mathrm{cd}} + t_{\mathrm{dd}} + t_{\mathrm{do}} + t_{\mathrm{oc}} + t_{\mathrm{cq}}
\approx \SI{10}{\micro\second}. \label{eq:estimated_com_time}
\end{equation}
This budget is dominated by controller--decoder--orchestrator data movement ($t_{\mathrm{cd}}, t_{\mathrm{do}}, t_{\mathrm{oc}}$), whose effective latency scales roughly as $\mathcal{O}(N_{\mathrm{ps}}/N_{\mathrm{ch}})$, where $N_{\mathrm{ps}}$ is the payload size and $N_{\mathrm{ch}}$ is the number of parallel channels. Because $N_{\mathrm{ch}}$ and link bandwidth are systems engineering choices, they can be further optimized (e.g., by widening channelization, batching, and overlapping transfers) in light of these measurements to satisfy the latency requirements of efficient FTQC.

\section{Reaction Time in Fault-Tolerant Quantum Computation}
\label{sec:reaction_time}

In this section, we discuss why the reaction time determines the overall speed of computation in our fault-tolerant microarchitecture, and present a method that combines the spatial- and temporal-window techniques~\cite{Skoric2023Parallel, FuhuiLin2025Spatially} to enable additional decoder parallelization that speeds up a given computation. Our analysis continues on from \cref{sec:surface-codes}, where it was explained that the $\pi/8$ gadget has one conditional operation: the measurement of the correction qubit in either the $X$- or the $Z$-basis. This decision is based on the outcomes of the $P \otimes Z$ lattice surgery measurement in the same $\pi/8$ gadget and the correction qubit measurement in a previous $\pi/8$ gadget if the two gadgets anti-commute. It can be assumed that, in the core processor, each adjacent pair of $\pi/8$ gadgets anti-commute, which is reasonable if the algorithm has been compiled efficiently. However, in the 15-to-1 magic state distillation protocol, all rotations are $Z$-type rotations, and thus commute with each other~\cite{Litinski2019Game}.

\begin{figure*}[tb]
    \centering
    \includegraphics[width=\linewidth]{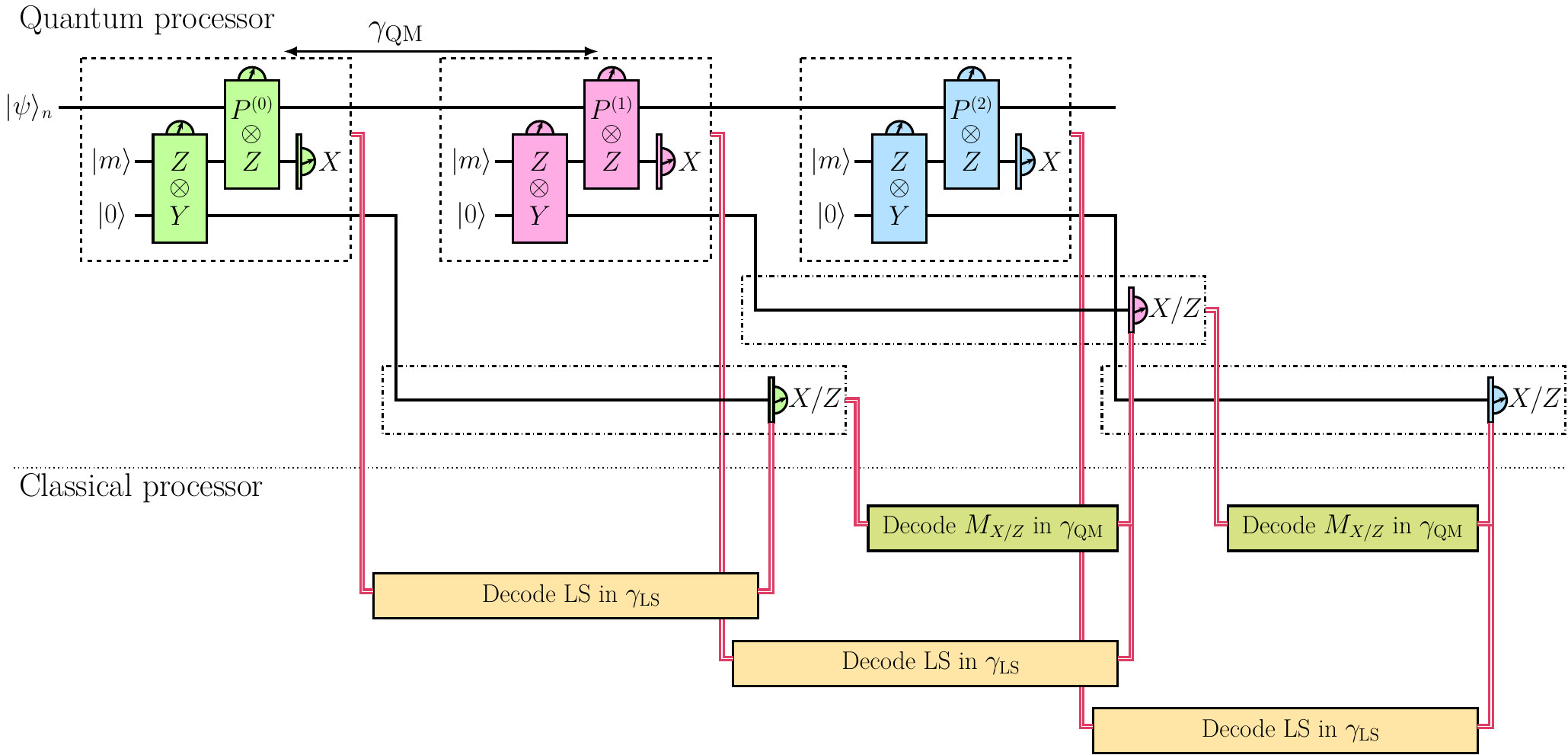}
    \caption{Implementation of a sequence of post-corrected $\pi/8$ Pauli rotations on a QPU. Each $\pi/8$ rotation gadget (indicated by a grey dashed line box) requires performing measurements involving logical data qubits in the joint state $|\psi\rangle$, a high-fidelity magic state $|m\rangle$ distilled in an RSF (not shown), and a correction qubit prepared in the state $|0\rangle$. The correction qubit in each rotation gadget is measured in either the $X$- or $Z$-basis, depending on the outcome of the logical measurements and on the outcome of the measurement of the previous correction qubit. In the classical processor, the received syndrome data is used to decode the outcomes of the measurements. The lattice surgery measurements involved in implementing each $\pi/8$ rotation are spatially large and have a reaction time of $\gamma_\text{LS}$. The correction qubits cannot be measured until their associated lattice surgery measurements have been decoded until which time they are stored in quantum memory. Hence, the decoding of each correction qubit measurement has a reaction time of $\gamma_\text{QM}$. The decoding of the logical measurements can be parallelized, but the decoding of the correction qubit measurements cannot be unless the $P^{(i)}$ commute.}
    \label{fig:pi8_diagram}
\end{figure*}

The correct value of a logical measurement is discovered once the decoder identifies all errors using the acquired syndrome data. The time it takes to send the syndromes of the $P\otimes Z$ lattice surgery in a gadget to the decoder, compute the value of the measurement, and send the measurement basis decision to the QPU is the first reaction time, labelled $\gamma_\text{LS}$. During this time, the correction qubit must remain stored in quantum memory, after which it may be measured. The second reaction time, labelled $\gamma_\text{QM}$, is the latency associated with determining the measurement basis of the next correction qubit based on the measurement outcome of the previous correction qubit. Even though both of these reactions are required for the same decision, they play a very different role in determining the schedule and the speed of the computation. In what follows, we first explain how the decoding of the joint measurement outcomes of the $P \otimes Z$ lattice surgeries in different gadgets can be parallelized. We then argue why, in the core processor, the decoding of the correction qubit measurement outcomes causes a slowdown in the quantum computation. This is illustrated in \cref{fig:pi8_diagram}. Finally, we show that this slowdown does not occur in the magic state factories.

The decoding of a lattice surgery is used to determine two key pieces of information: identifying error chains on the logical qubit patches and the outcome of the joint measurement ($m$ in \cref{eq:lattice_surgery_outcome}). The former can be ascertained once the previous logical operation has been decoded (in contrast, see the discussion in Ref.~\cite{Zhang2026Learning}). Na\"ively, the latter requires the same; however, by using the idea of temporal buffer regions, the value of $m$ can be determined independently of the decoding results of the previous logical operation. 

Recall that, in both parallel window approaches, the reason for adding buffer regions to the windows is to reliably decode the commit region. In these approaches, the buffers are added only in either the space or the time dimensions. In particular, in the space parallelization method the initial temporal boundaries of the windows must be specified, and are determined by the decoding result of the immediately preceding logical operation. If, however, a temporal buffer region is added at the beginning of each spatial window, and the commit region is extended temporally to overlap with the temporal buffer region of the next operation, then error chains that flip the joint measurement outcome can be determined without knowledge of the full error chain history. To summarize, in our proposed approach, decoding a $Y$ surgery will use windows of sizes
\begin{equation}
((2d, 2d, 3d), (2d, 1.5d, 3d), (d,d,3d)). \label{eq:ls-window-sizes}
\end{equation}
The initial temporal buffer will be resolved once the preceding logical operation has completed, at which point the Pauli frame for the logical qubits will be updated. Using this approach, the lattice surgery decoding does not slow down the computation, and $\gamma_\text{LS}$ determines only how long the correction qubit will be stored in memory before being measured. Note that the correction qubit is always stored in memory for one logical cycle during the $P \otimes Z$ surgery, so its total storage time is $\gamma_\text{LS} + \tau_\text{logical}$. The impact of this parallelization is felt through an increase in the number of required decoding units, the estimate of which is given in \cref{sec:decoder-how-many}.

In the core processor, the measurements of the correction qubits and the decoding of the corresponding measurement outcomes should be completed in sequence. This is because the decoding of the final two windows of the correction qubit before its measurement cannot be completed until the correction qubit has been measured in some basis, and this basis is decided only once the previous correction qubit's measurement outcome has been completely decoded. In the core processor, if subsequent gadgets anti-commute, then they should be executed with a spacing of $\gamma_\text{QM}$; otherwise, the correction qubits will have to be placed in storage, waiting to be decoded in sequence. For utility-scale quantum computing, this storage could be billions or trillions of logical qubits in size, which is infeasible. By executing rotation gates at the same rate as the reaction time, at most $\lceil \gamma_\text{LS}/\gamma_\text{QM}\rceil$ correction qubits will need to be stored in quantum memory in the core processor, assuming there is no parallelization of $T$-state injections. Hence, the total execution time of a quantum circuit is at least $\gamma_\text{QM}N_T$, where $N_T$ is the number of rotation gadgets in the circuit in the case of no parallelization, or the circuit depth in the case of parallelization. One reason this is an underestimate is that the RSF needs some warm-up time to begin to produce magic states. For the resource estimates made in this work, we assume that the execution of the quantum circuit is time-optimal.

In contrast to the circuit in the core processor, the circuits executed in the RSF are typically very short. For example, each \mbox{15-to-1} magic state distillation unit requires the implementation of only 11 rotations~\cite{Litinski2019Game}. These rotations can be performed sequentially every logical cycle time $\tau_\text{logical}$, producing one magic state every $11\tau_\text{logical}$. A newly produced magic state must remain stored until all associated correction qubits have been measured. Since all operations in the \mbox{15-to-1} distillation protocol commute, if a sufficient number of classical processing units is available to decode all measurement outcomes in parallel, then the waiting time for the stored magic state is bounded by $\gamma_\text{QM}$. Therefore, the steady-state output rate of each distillation unit is one magic state every $11\tau_\text{logical}$ assuming that the distillation process succeeds. Executing the rotation gadgets of the distillation unit as fast as possible requires an additional storage area of $\lceil \gamma_\text{QM}/\tau_\text{logical} \rceil$ patches in which both the correction qubits and the produced magic states are stored. This strategy prevents the multiplicative slowdown of magic state production when an RSF contains more than one level of distillation, and is hence necessary for the time-optimal execution of a quantum circuit.

We emphasize that in a realistic system, the reaction times increase as the patch distance increases, due to the increase in the number of syndromes to be transmitted and decoded. Consequently, the two reaction times vary across the different components of the microarchitecture. For example, even though in the microarchitecture correction qubits are stored everywhere for a time of $\gamma_\text{LS}$, this quantity is usually the smallest at the lowest RSF level, then subsequently larger at each higher level of the RSF, and the largest in the core processor. Therefore, the correction storage area's size per injection unit is typically smaller at the lower levels of the RSF than at the higher levels.

\section{Effects of Reaction Time on Quantum Resource Estimates}
\label{sec:results-resource-estimates}

As discussed in the previous section, for a nonzero reaction time, the time-optimal runtime of a quantum circuit is generally approximated by $\gamma_\text{QM}N_T$. However, the space cost must be calculated carefully when optimizing the microarchitecture, as described in \cref{sec:ftqc-assembly}. The reaction time affects this optimization problem in three ways.

First, as the two reaction times increase, the logical qubits must be stored in memory for increasing amounts of time resulting in the further accumulation of logical error, which must be counteracted by increasing the code distances used in both the core processor and the RSF, so as to keep the total logical error below the specified error budget. Second, the total size of the correction qubit storage and the magic state storage in the RSF of size $\lceil \gamma_\text{QM}/\tau_\text{logical} \rceil$ increases proportionally with the reaction time, which may also increase the code distances used in the RSF to prevent the error budget from being exceeded. Third, as the reaction time increases, the rate at which magic states must be provided to the core processor decreases, resulting in a smaller RSF. 

Hence, to study the impacts of the reaction time on quantum resource estimation, in \cref{sec:logical-error-rate-pi8}, we present a quantitative analysis of how changes in the reaction times affect the logical error rates of a computation. Then, in \cref{sec:resource-estimate-results}, we discuss our results on how the physical qubit count varies with the reaction time for two quantum algorithms.

\subsection{Impact of Reaction Times on the Logical Error Rates of a Quantum Computation}
\label{sec:logical-error-rate-pi8}

We now describe how the two reaction times contribute to the logical error rate in a computation. To this end, we construct a constitutive model of the logical error rate of the $\pi/8$ rotation gadget, based on logical error rate models of quantum memory and lattice surgery. These logical error models are derived from the geometry of space--time diagrams of the quantum error correcting protocols, where the lengths and the areas of various surfaces determine the coefficients in these models (see Fig.~8 in Ref.~\cite{Silva2025Optimizing}).

For quantum memory on a square rotated surface code patch, the logical error rate is
\begin{equation}
    \text{Pr}_\text{QM}(d, r) = \mu_S dr \Lambda_S^{-(d+1)/2}, \label{eq:ler_memory}
\end{equation}
where $d$ is the code distance, $r$ is the number of syndrome measurement rounds, $\Lambda_S$ is the error suppression rate for spatial logical errors, and $\mu_S$ is a pre-factor for spatial error rates~\cite{Fowler2012Surface, Chamberland2022Universal, Kelly2015State, Acharya2025Quantum}. In our logical microarchitecture, we set $r=d$, because the optimal number of rounds for logical operations in the surface code is generally believed to be equal to the code distance.

In our logical microarchitecture, lattice surgeries are typically performed on multiple logical qubits connected by a bus consisting of multiple patches. We present a model of the logical error rates of such surgeries. Suppose a lattice surgery of type $Q=X^{\otimes K}$ is performed on $K$ logical qubits encoded in surface code patches of distance $d$ using a bus consisting of $B$ logical patches of equal distance. Then, carefully accounting for the geometry of the space--time model associated with this lattice surgery, the logical error rate of the surgery is~\cite{Mohseni2026How,Silva2025Optimizing,Chamberland2022Universal}
\begin{equation}
    \text{Pr}_\text{LS}(d, r, K, B) = \mu_X (K + B)d r\Lambda_X^{-(d+1)/2}  + \mu_Z Kd\Lambda_Z^{-(d+1)/2} + \mu_T B d^2 \Lambda_T^{-(r+1)/2}, \label{eq:ler_ls_full}
\end{equation}
where $r$ is the number of syndrome measurement rounds used in implementing the merging operations.
The first term here accounts for the logical $X$ errors, which can occur anywhere on the logical qubits or the bus. However, logical $Z$ errors can occur only on the logical qubits, and only just prior to merging or just after splitting. Hence, the second term has a smaller coefficient. The third term represents the time-like errors that can occur only on the bus patches and thus cause the value of the joint measurement to be incorrectly inferred. We assume an unbiased noise model, with identical error suppression for the $X$ and the $Z$ logical errors. Hence, we can set $\mu_S = \mu_X = \mu_Z$ and $\Lambda_S=\Lambda_X=\Lambda_Z$ and obtain the simplified expression
\begin{equation}
    \text{Pr}_\text{LS}(d, r, K, B) = \mu_S[(K + B)d r + Kd]\Lambda_S^{-(d+1)/2} + \mu_T B d^2  \Lambda_T^{-(r+1)/2}, \label{eq:ler_ls_simplified}
\end{equation}
which, due to the symmetry between $X$ and $Z$, is applicable to lattice surgeries of any weight-$K$ Pauli operator $Q$.

With these models in hand, we turn to the analysis of the post-corrected $\pi/8$ rotation circuit shown in \cref{fig:pi8_gadget}. The first lattice surgery creates a two-qubit resource state. In our logical microarchitecture, the correction qubit is encoded in a $d \times 2d$ rectangular code patch. This allows $Y$ lattice surgeries to be performed in a twist-based architecture~\cite{Litinski2018Lattice}. A two-tile bus is needed to connect the $Z$ logical operator of the magic state to the $X$ and $Z$ logical operators of the correction qubit. Hence, for this surgery, $K=3$ and $B=2$. The second surgery involves one of the qubits of the joint two-qubit resource state as well as $K$ logical data qubits. In our simulations, we determine the average number of logical data qubits $K_\text{avg}$ in the compiled circuit, and estimate the average number of bus tiles $B_\text{avg}$ used in each surgery. Hence, for this surgery, $K = K_\text{avg} + 1$ and $B = B_\text{avg}$.

The reaction times impact the overall logical error rate of the $\pi/8$ rotation because they result in a need for additional storage of qubits in quantum memory. The correction qubit is stored in memory for $\gamma_\text{LS} + \tau_\text{logical}$ and the computation qubits are stored in memory for $\gamma_\text{QM}$, yielding the total logical error rate of the gadget
\begin{equation}
    \text{Pr}_{\frac{\pi}{8}}(d) = \text{Pr}_\text{LS}(d, d, 3, 2)  + \text{Pr}_\text{LS}(d, d, K_\text{avg} + 1, B_\text{avg}) + \frac{K_\text{avg}\gamma_\text{QM} + \gamma_\text{LS} + \tau_\text{logical}}{\tau_\text{logical}} \text{Pr}_\text{QM}(d, d),\label{eq:ler_pi8_with_d}
\end{equation}
where the division by $\tau_\text{logical}$ converts the wall-clock time into a time expressed in terms of the number of logical cycles, because $\text{Pr}_\text{QM}$ is the error of a single logical cycle. This model can be incorporated into the error budget calculations used in our resource estimation analyses~\cite{Silva2025Optimizing}, discussed in \cref{sec:resource-estimate-results}. For these analyses, we use the noise model defined in Appendix~\ref{sec:noise-model} to obtain the error rates of quantum memory and lattice surgery, which are presented in Appendix~\ref{sec:ftqc-performance}. Note that our noise model implies that a stabilizer measurement round takes only $\SI{0.35}{\micro\second}$. This does not include the time taken for error suppression techniques like dynamic decoupling or leakage removal, which need to be employed on realistic quantum processors, and result in a stabilization time of $\sim$$\SI{1}{\micro\second}$~\cite{Miao2023Overcoming, Acharya2025Quantum}.

\subsection{Impact of Reaction Time on Physical Qubit Requirements}
\label{sec:resource-estimate-results}

\begin{figure*}[tb]
    \centering
    \begin{subfigure}[b]{0.48\textwidth}
        \centering
        \includegraphics[width=\textwidth]{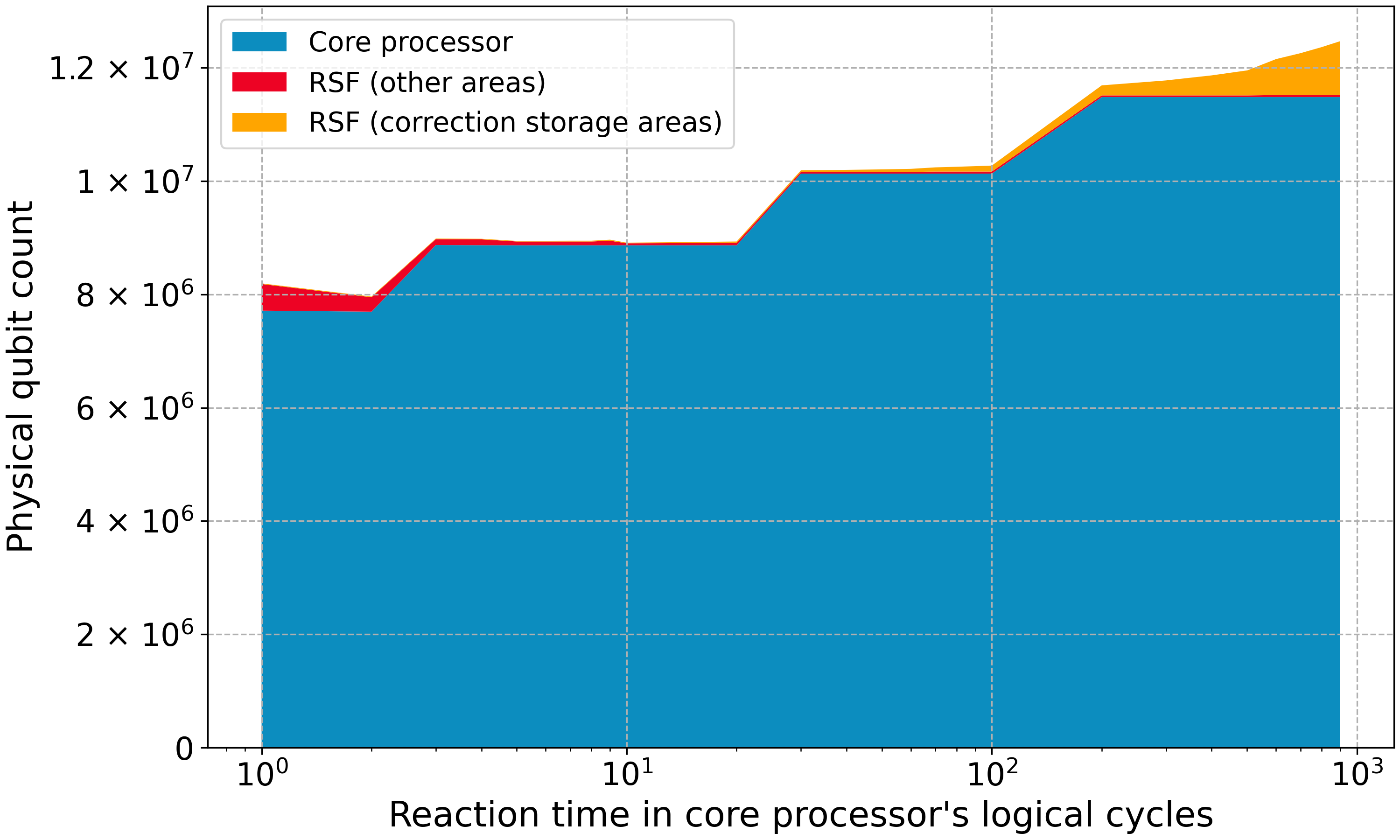}
        \caption{Quantum resource estimation for a circuit with a low $T$ count and a high logical qubit count}
        \label{fig:low_T_high_qubit}
    \end{subfigure}
    \hfill
    \begin{subfigure}[b]{0.48\textwidth}
        \centering
        \includegraphics[width=\textwidth]{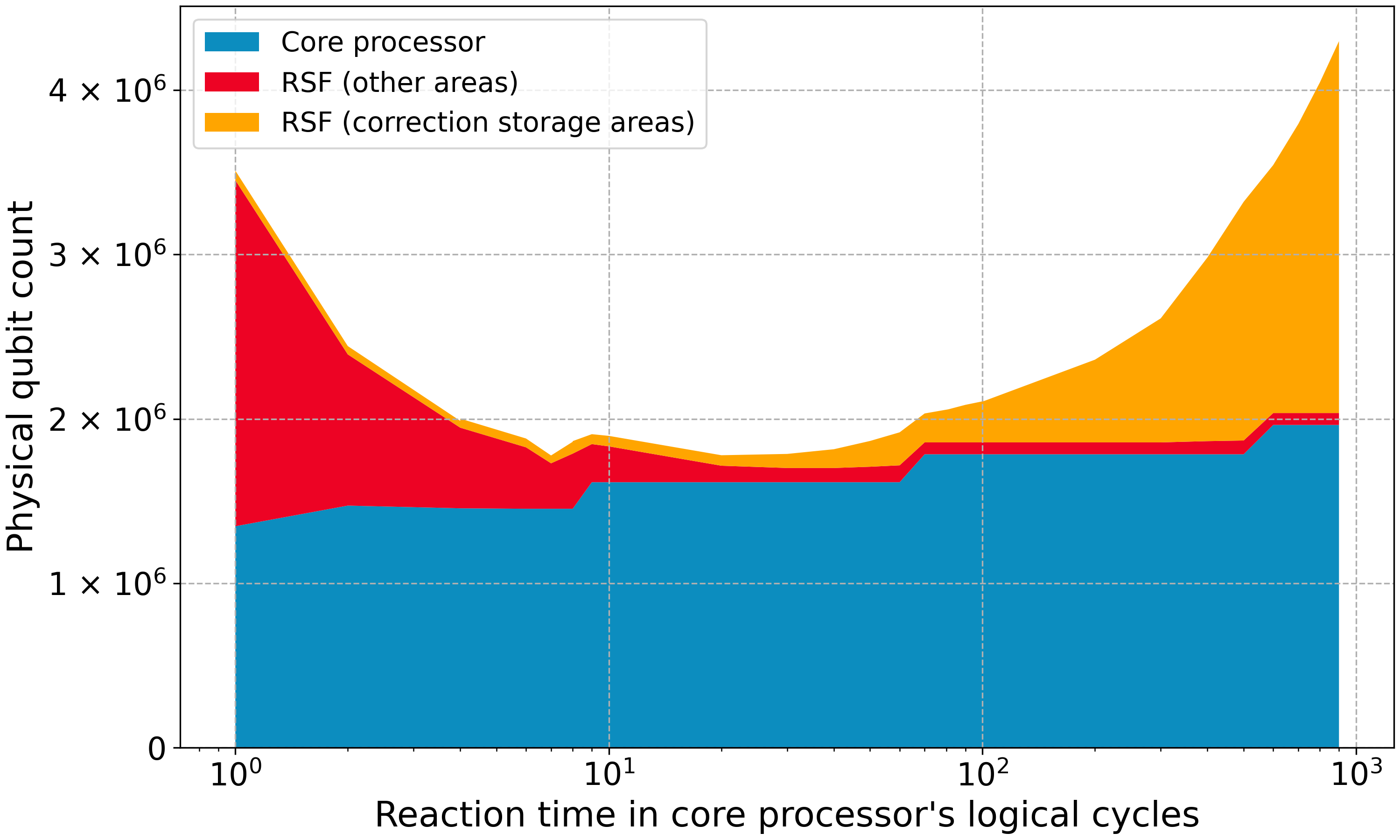}
        \caption{Quantum resource estimation for a circuit with a high $T$ count and a low logical qubit count}
        \label{fig:high_T_low_qubit}
    \end{subfigure}
    \caption{Physical qubit footprint versus reaction time, measured in logical cycles of the core processor, broken down by microarchitecture area. Estimates are provided for two contrasting utility-scale quantum algorithms: (a) ground-state energy estimation targeting a precision of \mbox{0.005$L^2$} for a 2D Fermi--Hubbard model on an $L \times L$ lattice, using a circuit for lattice size $L=32$ generated using Campbell's plaquette Trotterization approach~\cite{Campbell2022Early}, and which requires 2562 logical qubits and $4 \times 10^6$ $T$ gates; and (b) a single-shot dynamic-circuit implementation of the quantum eigenvalue transform for NMR spectral prediction of the $\alpha$-conotoxin macromolecule~\cite{Elenewski2026Prospects}, requiring 241 logical qubits and $5.11 \times 10^{11}$ $T$ gates. For each algorithm, the figure shows the architecture chosen on the time-optimal branch of the space--time Pareto frontier generated using the TopQAD software suite~\cite{1qbit2024topqad}. }
    \label{fig:qre_per_area}
\end{figure*}

We now analyze the impact of reaction time on the space costs of FTQC. To this end, we assemble two utility-scale quantum algorithm circuits onto the LMPSC microarchitecture. The Fermi--Hubbard plaquette Trotterization circuit~\cite{Campbell2022Early} is characterized by a larger logical qubit count and a relatively lower $T$ count, whereas the circuit for NMR spectral prediction for $\alpha$-conotoxin~\cite{Elenewski2026Prospects} represents the opposite regime, with a higher $T$ count and fewer logical qubits. For both cases, we choose the time-optimal architectures on the space--time Pareto frontier, and plot in \cref{fig:qre_per_area} the resultant space costs as a function of reaction time, assuming $\gamma=\gamma_\text{QM} = \gamma_\text{LS}$. The total space costs are broken down into three areas of the microarchitecture: the core processor, the RSF correction storage area, and the remaining areas of the RSF. The trends in each curve reveal which areas of the microarchitecture dominate the space overhead in different reaction time regimes.

We observe from the plots a clear trade-off between the RSF correction storage and remaining areas. Doubling the reaction time approximately halves the number of distillation units required in the RSF, thereby reducing the footprint of the RSF area used for distillation. This occurs because a longer reaction time lowers the required magic state throughput. However, each additional logical cycle of reaction time also requires one extra logical patch for correction qubit storage to buffer delayed decoding feedback. Consequently, doubling the reaction time doubles the storage overhead. Within the utility-scale range tested, the RSF code distances vary from 23 in the single-level RSF for the shorter circuit (see \cref{fig:low_T_high_qubit}) to 13 and 33 in the two-level RSF for the longer circuit (see \cref{fig:high_T_low_qubit}). This results in an additional space overhead of roughly 1000--2500 physical qubits per extra logical cycle of reaction time. For the hardware model analyzed, this corresponds to an extra 100,000--250,000 physical qubits when decoder latencies are on the order of one millisecond, that is, 100 logical cycles of the core processor.

Another notable effect we observe is a gradual growth of the core processor footprint with increasing reaction time, which can be easily explained to be due to the increased protection requirements as the circuit execution time lengthens. In particular, we find that the code distance in the core processor increases by 2 for each reaction time increase of approximately a factor of $\Lambda$. In our simulations, where $\Lambda=9.3$ (see Appendix \ref{sec:ftqc-performance}), we find that, as the reaction time is changed from $\gamma=1$ logical cycle to $\gamma=100$ logical cycles, the core processor's distance increases from 27 to 31 for the shorter circuit, and from 35 to 39 for the longer circuit. These distance increases correspond to roughly $32\%$ and $24\%$ increases in the physical qubit counts, respectively. For the circuits tested in the 200--2000 logical qubit range, this overhead corresponds to an additional approximate increase of 300,000--1,750,000 physical qubits.

The dominant contributor to the total space cost depends on a circuit's characteristics. For circuits requiring more computation qubits (see \cref{fig:low_T_high_qubit}), the core processor area dominates. In contrast, for circuits with a relatively higher $T$ count (see \cref{fig:high_T_low_qubit}), the RSF area becomes the main driver of space cost. In this case, at short reaction times, the large number of parallel distillation units needed to sustain the required injection rate dominates the space overhead, and at sufficiently large reaction times, correction storage patches become the primary contributor.

These results show that developing faster decoders or communication links can decrease the distances required for utility-scale computations. Since decoding speed is typically a function of distance, this can set up a feedback loop of improvement. However, for circuits with a high $T$ count relative to the number of computation qubits, decreasing the reaction time is counterproductive from a physical qubit count perspective. This problem can be alleviated by the development of either smaller magic state preparation units and distillation factory units, or ones with better error rates which can reduce the number of distillation levels required.

\section{Decoder Requirements}
\label{sec:decoder-requirements}

In this section, we determine how fast individual decoding units need to be, and how many are needed for utility-scale quantum computing. In \cref{sec:decoder-speed}, we present a model for the reaction time, and use the model to analyze how fast decoding units need to be. In \cref{sec:decoder-how-many}, we provide an estimate as to the number of decoders needed for a 10M-qubit QPU.

\subsection{Speed of Decoding Units}
\label{sec:decoder-speed}
The reaction time is the sum of the decoding time and the communication latencies in the system, the latter of which are estimated in \cref{sec:communication-latency}. The decoding time required by a single decoding unit depends on the size of the decoding graph passed to it. The decoding time per stabilization round, or the decoding speed, is typically modelled as 
\begin{equation}
\tau_\text{decoding}(N) = \alpha N^\beta, \label{eq:decoder-space-model}
\end{equation}
where $N$ is the number of detectors associated with a single spatial layer of the decoding graph. The total decoding time is given by $r\tau_\text{decoding}(N)$, where $r$ is the number of temporal layers in the decoding graph. This model has been used to report the decoding time of a $d\times d$ memory patch for a number of decoders, some of which are listed in \cref{tab:decoder-scaling}.

\begin{table}[!tb]
\centering
\begin{tabular}{lccc}
\toprule
Decoder & $\tau_\text{decoding}(N)$ & Reference \\
\midrule
Collision Cluster on ASIC & \multirow[t]{2}{*}{$5.53 \times 10^{-11} N^{1.34}$} & \multirow[t]{2}{*}{Table~2 in Ref.~\cite{Barber2025Realtime}}\\
\quad (CC-ASIC) & & \\
Collision Cluster on FPGA & \multirow[t]{2}{*}{$2.85 \times 10^{-10} N^{1.20}$} & \multirow[t]{2}{*}{Fig.~4 in Ref.~\cite{Barber2025Realtime}} \\
\quad (CC-FPGA) & & \\
AlphaQubit & $4.80 \times 10^{-6} N^{0.50}$ & Extended Data Fig.~7 in Ref.~\cite{Bausch2024Learning} \\
PyMatching at $p=0.1\%$ & $5.92 \times 10^{-9} N^{1.17}$ & Extended Data Fig.~7 in Ref.~\cite{Bausch2024Learning} \\
\bottomrule
\end{tabular}
\caption{Decoding-time scaling for four surface code decoders under the uniform depolarizing model with physical error rate $p=0.1\%$. The values of $\alpha$ and $\beta$ in \cref{eq:decoder-space-model} for each decoder are numerically estimated by extracting data points from the indicated figures and table.}
\label{tab:decoder-scaling}
\end{table}

Using this model of the speed of a decoding unit, and recalling that temporal parallelization requires windows of size $3d$ (see list \ref{eq:ls-window-sizes}), the reaction time of a square quantum memory is
\begin{align}
    \gamma_\text{QM} &= 3d\tau_\text{decoding}(d^2) + t_\text{dd} + 3d\tau_\text{decoding}(d^2) + t_\text{qc} + t_\text{cd} + t_\text{do} + t_\text{oc} + t_\text{cq} \nonumber\\
    &= 6d\tau_\text{decoding}(d^2) + t_\text{dd} + t_\text{qc} + t_\text{cd} + t_\text{do} + t_\text{oc} + t_\text{cq} \nonumber \\
    &= 6d\tau_\text{decoding}(d^2) + \tau_\text{com}, \label{eq:mem_reaction_time_model}
\end{align}
where the total communication time is
\begin{equation}
    \tau_\text{com} = t_\text{dd} + t_\text{qc} + t_\text{cd} + t_\text{do} + t_\text{oc} + t_\text{cq},
\end{equation}
which is estimated to be $\tau_\text{com} \approx \SI{10}{\micro\second}$ in \cref{eq:estimated_com_time}. Many published quantum resource estimates assume a reaction time of $\SI{10}{\micro\second}$~\cite{Gidney2021How, Lee2021Even, Litinski2022Active, Gidney2025How}, but our results show that this budget is consumed by the communication latency alone. For comparison, using the window sizes in list~\ref{eq:ls-window-sizes}, the $Y$-type lattice surgery reaction time is
\begin{align}
    \gamma_\text{LS} &= 3d\tau_\text{decoding}(4d^2) + t_\text{dd} + 3d\tau_\text{decoding}(3d^2) + t_\text{dd} + 3d\tau_\text{decoding}(d^2) \nonumber \\&\qquad+ t_\text{qc} + t_\text{cd} + t_\text{do} + t_\text{oc} + t_\text{cq} \nonumber \\
    &= 3d(4^\beta + 3^\beta + 1)\tau_\text{decoding}(d^2) + 2t_\text{dd} + t_\text{qc} + t_\text{cd} + t_\text{do} + t_\text{oc} + t_\text{cq}, \label{eq:ls_reaction_time_model}
\end{align}
where the communication time is slightly higher due to the use of three layers of decoding. 

In the case of time-optimal compilation, the memory's reaction time is related to the runtime $t_\text{circuit}$ of a compiled quantum circuit with $N_T$ magic state injections ($T$ count) via
\begin{equation}
    t_\text{circuit} \approx \gamma_\text{QM}N_T.
\end{equation}
This equation and \cref{eq:mem_reaction_time_model} can be combined to yield the relation
\begin{equation}
    \tau_\text{decoding}(d^2) = \frac{t_\text{circuit}/N_T - \tau_\text{com}}{6d}
\end{equation}
between the decoding speed of a square memory patch, the total circuit execution time, the $T$ count, and the distance of the core processor. This equation enables a uniform comparison of the capabilities of different decoders to be made.

\begin{figure}[tb]
    \centering
    \includegraphics{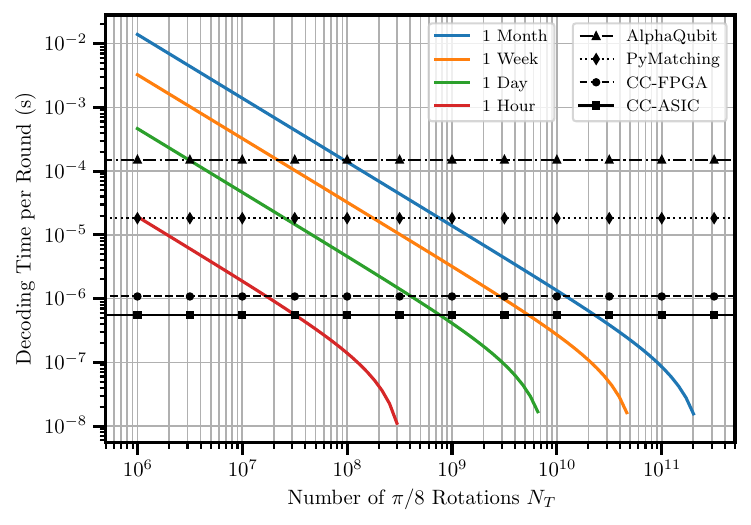}
    \caption{Decoding time of a square memory patch $\tau(d^2)$ versus the number of magic state injections in a quantum circuit for four circuit execution times, assuming that the core processor operates at $d=31$. Also plotted are the decoding times of the decoders listed in \cref{tab:decoder-scaling} for $d=31$. The fastest decoder (i.e., the CC-ASIC decoder) can execute circuits with $N_T \le 3\times 10^7$ in an hour, and those with $N_T \le 2\times 10^{10}$ in a month. Note that the communication latency $\tau_\text{com}$ will begin to be a bottleneck below a decoding time of $10^{-7}$.}
    \label{fig:decoder-speed-T}
\end{figure}

\begin{figure}[tb]
    \centering
    \includegraphics{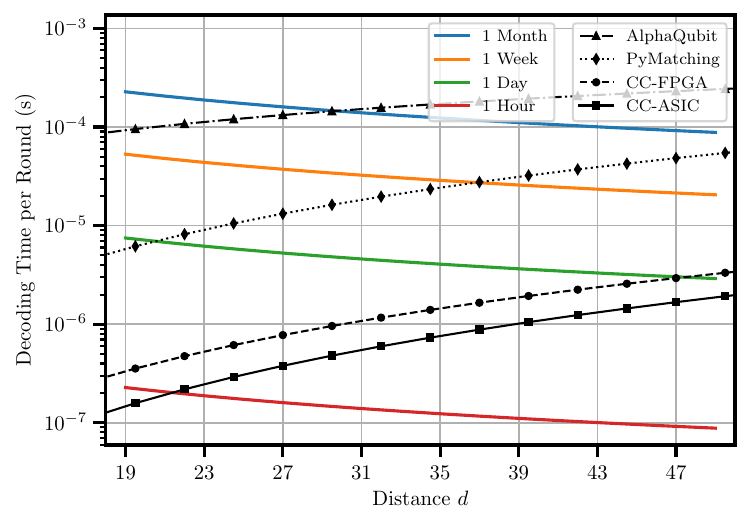}
    \caption{Decoding speed of a square memory patch $\tau(d^2)$ versus the distance of the core processor for four circuit execution times, assuming a circuit $N_T$ count of $10^8$. Also plotted as a function of distance are the decoding speeds of the decoders listed in \cref{tab:decoder-scaling}. The decoding speed is not very sensitive to the distance at this $N_T$ count.}
    \label{fig:decoder-speed-d}
\end{figure}

\Cref{fig:decoder-speed-T} plots, for $d=31$, the decoding speed as a function of $N_T$ for four circuit execution times $t_\text{circuit}$. It shows that, given a fixed circuit execution time, an improvement in decoding speed by an order of magnitude can yield improvements in $N_T$ by roughly an order of magnitude. However, this advantage does not last indefinitely because, as decoders become faster, the communication latency begins to dominate the reaction time. If communication latencies cannot be reduced further, then other techniques will be needed to help make utility-scale quantum computing viable, such as compilation methods or algorithmic improvements that significantly reduce the $T$ count. Another way to mitigate the impact of large reaction times is to improve the physical error rates of the QPU. This is explored in \cref{fig:decoder-speed-d}, which plots, for $N_T = 10^8$, the decoding speed as a function of distance. Even though the two quantities are inversely related, halving the distance manages to improve decoding speed by only about a third of an order of magnitude. The expected improvements in errors rates for superconducting qubits in the medium term will not result in such drastic improvements in distance.

\subsection{Number and Size of Decoding Units}
\label{sec:decoder-how-many}

The minimal requirements for decoding space and time for a quantum computer are substantial, and the parallelization of lattice surgery decoding, discussed in \cref{sec:reaction_time}, adds further costs. Therefore, we estimate the number and size of the decoding units required to keep up with the syndrome throughput, ignoring any over- or under-estimation that could arise due to decoder scheduling constraints~\cite{Maurya2025Managing}. For simplicity, we restrict our analysis to the core processor, and ignore the RSF. In the core processor, only two logical operations occur. All computation qubits must be stored in quantum memory, unless they are involved in the $P \otimes Z$ lattice surgery of the rotation gadget, which happens every $\lceil\gamma_\text{QM}/\tau_\text{logical}\rceil$ logical cycles. We make the worst-case assumption that the circuit has been compiled such that there is at least one multi-qubit operator $P$ in the circuit that acts on all $C$ computation qubits and uses all $B$ bus patches. Note that in the LMPSC microarchitecture family $B \approx C$.

When the computation qubits are in memory, the number of decoders required can be estimated by Little's law to yield a formula similar to Eq.~(2) in Ref.~\cite{Skoric2023Parallel}. In the steady state, two layers of decoding have latency $2\tau_w + t_\text{dd}$, where $\tau_w$ is the time required to decode a single window. As computation qubits are stored in pairs, the size of a window is $(2d, d, 3d)$, which yields $\tau_w = 3d\tau_\text{decoding}(2d^2)$. During the execution of two layers, $4d$ rounds are committed, which yields a throughput of $1/4\tau_\text{logical}$. Hence, the number of decoders required for a single patch is
\begin{equation}
    k_{\text{QM},1} = \left\lceil\frac{2\tau_w + t_\text{dd}}{4\tau_\text{logical}}\right\rceil,
\end{equation}
and for $C$ computation qubits stored in pairs, the equivalent number is
\begin{equation}
    k_\text{QM} = \left\lceil\frac{C[6d\tau_\text{decoding}(2d^2) + t_\text{dd}]}{8\tau_\text{logical}}\right\rceil.
\end{equation}

For decoding a single lattice surgery, the parallelization algorithm has three layers, in each of which one-third of the cross-sectional area is committed. Hence, the number of parallel decoders needed is one-third of the number of patches in the surgery. According to our assumption above, there is a lattice surgery that involves $C+B\approx 2C$ patches, which requires 
\begin{equation}
    k_{\text{LS}, 1} = \left\lceil\frac{2C}{3}\right\rceil
\end{equation}
parallel decoders. The first layer involves the largest problems of size $(2d, 2d, 3d)$ (see list~\ref{eq:ls-window-sizes}), which sets the minimum size for the decoding units. Further, as discussed in \cref{sec:reaction_time}, the decoding of lattice surgery can be parallelized, and $\lceil\gamma_\text{LS}/\gamma_\text{QM}\rceil$ such decodings can occur in parallel. Hence, for the lattice surgery decoding, we estimate that
\begin{equation}
    k_\text{LS} = \left\lceil\frac{2C}{3}\right\rceil \times \left\lceil\frac{\gamma_\text{LS}}{\gamma_\text{QM}}\right\rceil
\end{equation}
parallel decoders are needed. Given that $\gamma_\text{LS} > \gamma_\text{QM}$ for most decoders, both the memory and lattice surgery decoding tasks happen in parallel; hence, the total number of decoding units needed is
\begin{equation}
    k = \left\lceil\frac{C[6d\tau_\text{decoding}(2d^2) + t_\text{dd}]}{8\tau_\text{logical}}\right\rceil + \left\lceil\frac{2C\gamma_\text{LS}}{3\gamma_\text{QM}}\right\rceil.
\end{equation}
We estimate this number assuming we have a core processor that operates at $d=31$ and occupies $90\%$ of a 10M-qubit QPU, the CC-ASIC decoding model is used, and communication latencies in the system are as provided in \cref{tab:data-transfer-time}. This results in $k\approx 11,700$. Adding a $10\%$ additional cost for the RSF, we estimate that the decoder cluster will need about $\sim$$13,000$ decoding units, which is a fairly substantial cost. This number can potentially be improved with alternative codes, more-efficient FTQC architectures, or improved decoder parallelization algorithms.

We conclude this section by computing the maximum space required for storing the syndromes in the decoder cluster's memory. The syndromes associated with the lattice surgery must be stored in the decoder cluster's memory for $2d(4^\beta + 3^\beta + 1)\tau_\text{decoding}(d^2) + 2t_\text{dd}$ (see \cref{eq:ls_reaction_time_model}). Assuming that all 5M syndrome qubits in a 10M-qubit QPU are stored for this long, the memory required is $\sim$$\SI{1800}{MB}$. This is an overestimate because the syndromes of each layer can be discarded as soon as the associated decoding tasks are complete, and an underestimate because it ignores the lattice surgery parallelization overhead.

\section{Conclusion}
\label{sec:conclusion}

In this paper, we have presented the results of our analysis of the performance of a quantum computer with respect to the speed of its decoding units and the latencies of its classical control electronics. Since the LMPSC logical microarchitecture (see \cref{fig:architecture_with_correction_storage}) relies on post-corrected magic state injection, the reaction time associated with quantum memories pertaining to correction qubits of the injection gadget is the main bottleneck for the speed of a quantum computation. Based on this insight, we built a model for the additional logical error that accumulates while the qubits are idling between subsequent magic state injections. Using this model, we built a procedure to find the time-optimal LMPSC microarchitecture for a quantum circuit. We conducted quantitative numerical studies of physical qubit count as a function of the reaction time for two quantum algorithms: ground-state energy estimation of a 2D Fermi--Hubbard model based on quantum phase estimation using Campbell's plaquette Trotterization approach~\cite{Campbell2022Early}; and NMR spectral prediction for a macromolecule based on the quantum eigenvalue transform~\cite{Elenewski2026Prospects}. These studies gave rise to the following three key insights. First, improving the reaction time by a factor of two allows for doubling the size of the resource state factories because magic states can be consumed faster in the core processor. Second, the same factor-of-two improvement in the reaction time also halves the space required to store the correction qubits in the resource state factories. Third, a decrease in reaction time by a factor of $\Lambda$ (which is the error suppression factor for quantum memory) results in a decrease in the core processor's code distance by 2. Our results highlight the trade-off between the size of the quantum processor and its physical error rates, given a fixed decoding speed.

We also designed a system architecture for the quantum execution environment of the computer, and experimentally estimated the latencies of transferring information between its components. We combined these communication latencies with the decoding speed of parallel space- and time-window decoders to build models for the reaction times. Our estimates of reaction times are much greater than the $\SI{10}{\micro\second}$ figure typically assumed in the literature. Using the reaction time models, we established, given a fixed circuit execution time, the trade-off curves between decoder speed and both the $T$ count and the distance of the code. Our results show that the fastest decoders currently available will need weeks or months to execute longer utility-scale quantum circuits. However, we found that an order of magnitude improvement in decoding speed can make circuits with an order of magnitude higher $T$ count accessible in the same execution time. Finally, we estimated that the execution environment of a 10M-qubit computer will require a thousand-channel high-speed communication link between the orchestrator, the controller, and the decoder cluster, with the decoder cluster consisting of $\sim$13k nodes, each capable of decoding independent patches of distance $\sim$30--60.

It is important to reiterate that these numbers are based on a syndrome extraction time of $\SI{0.35}{\micro\second}$, while on superconducting hardware, the relevant time is about $\SI{1}{\micro\second}$. Since the number of decoders is inversely proportional to this time, a more accurate estimate of the number of decoders will be closer to 5k. Our analysis also ignores that the decoder latency has significant variance because some decoding tasks are considerably harder that others, and that decoding requirements fluctuate substantially during the course of the algorithm's execution~\cite{Maurya2025Managing, Maurya2025FPGAtailored}. This inconstancy must be tackled by carefully designed scheduling algorithms for the decoding tasks, where the particular scheduling algorithm chosen will further refine our estimates.

The intention of our work has been to ascertain the magnitude of the improvements needed in decoding speeds and communication electronics, as well as other aspects of a quantum computer, to make FTQC a reality. Our results suggest that decoders and/or control electronics need to be significantly faster than those currently available to execute utility-scale circuits. However, improved compilation to reduce the $T$ count of algorithms or improved error rates to reduce the distance of the core processor can also expand the utility landscape. Another promising avenue for future development is the design of codes with better rates and distances than the surface code. Our results also show that methods of magic state preparation and distillation with lower space costs or error rates can significantly reduce the total physical qubit count.

\section*{Acknowledgements}

We thank our editor, Marko~Bucyk, for his careful review and editing of the manuscript. We thank Alan Ho, John Martinis, Namit Anand, Michael~E.~Beverland, Satvik Maurya, Matthias Troyer, and Boyan Torosov for useful conversations. The authors acknowledge the financial support of Pacific Economic Development Canada (PacifiCan) under project number PC0008525, and DARPA's Quantum Benchmarking Initiative (QBI) under project number \mbox{HR0011-25-9-0116}. G.~A.~D. is grateful for the support of Mitacs. P.~R. acknowledges the financial support of Innovation, Science and Economic Development Canada (ISED), the Province of Ontario through the Ministry of Colleges and Universities, and the Perimeter Institute for Theoretical Physics.

\begin{appendices}
\renewcommand{\thesection}{\Alph{section}}

\section{The Rotation Gadget}
\label{sec:rotation-gadget}
We further explain the operations of the $\pi/8$ rotation gadget presented in \cref{sec:surface-codes} and \cref{fig:pi8_gadget}~\cite{Litinski2019Game}. This gadget employs the magic state
\begin{equation}
\ket{m} = \cos(\pi/8)\ket{+} - i\sin(\pi/8)\ket{-}.
\end{equation}
Let us label the four measurements in the gadget as $m_i \in \{-1, +1\}$ for $i = 1,2,3,4$. The state of the system after the first three measurements is
\begin{align}
    & \left[\frac{(\bra{+}_1(1+m_3) + \bra{-}(1-m_3))}{\sqrt{2}}\right]\left(\frac{I + m_2P_0Z_1}{\sqrt{2}}\right)\left(\frac{I + m_1Z_1Y_2}{\sqrt{2}}\right)\ket{\psi}\ket{m}\ket{-})\ket{0}, \\
    &\quad= e^{-im_2\frac{\pi}{8}P_0}\left[ (m_2P_0)^{\frac{1-m_3}{2}}\ket{\psi}\ket{0} + im_1(m_2P_0)^{\frac{1+m_3}{2}}\ket{\psi}\ket{1} \right],
\end{align}
where $P_0$ is an $n$-qubit operator. Subsequently, measuring in the $Z$ basis yields the state
\begin{equation}
 P^{\frac{1-m_3m_4}{2}}e^{-im_2\frac{\pi}{8}P}\ket{\psi},
\end{equation}
and measuring in the $X$ basis yields the state
\begin{equation}
 P^{\frac{1-m_1m_3m_4}{2}}e^{im_2\frac{\pi}{8}P}\ket{\psi},
\end{equation}
where we drop the qubit index subscript on $P$. Hence, measuring in the $Z$ basis rotates the state by $m_2\pi/8$ and measuring in the $X$ basis rotates it by $-m_2\pi/8$, and in both cases there is a $50\%$ chance that an additional factor of $P$ is applied to $\ket{\psi}$ based on the outcomes $m_1$, $m_3$, and $m_4$. Since $P$ is a Pauli operator, this additional factor can be multiplied with the logical Pauli frame and corrected in software. The decision of the  basis in which to measure the correction qubit depends on the state $F$ of the Pauli frame, and the outcome $m_2$. Suppose that, at some point the state of the computation qubits is $F\ket{\phi}$, and we intend to rotate this state by $\pi/8$ around Pauli $P$. Note that if $F$ and $P$ anti-commute, then
\begin{equation}
    e^{\pm i\frac{\pi}{8}P}F = Fe^{\mp i\frac{\pi}{8}P},
\end{equation}
but this flip in the sign of the rotation does not occur if $F$ and $P$ commute. Hence, we must choose the measurement basis such that, if $F$ and $P$ commute $m_2\pi/8 = \pi/8$, and if $F$ and $P$ anti-commute $m_2\pi/8 = -\pi/8$. For example, in the latter case, the application of the rotation gadget will yield the intended state
\begin{equation}
    P^{\frac{1-\bar{m}}{2}}e^{i\frac{\pi}{8}P}F\ket{\phi} = \left(P^{\frac{1-\bar{m}}{2}}F\right)e^{-i\frac{\pi}{8}P}\ket{\phi},
\end{equation}
where $\bar{m}$ is either $m_3m_4$ or $m_1m_3m_4$, depending on the measurement basis, and $\left(P^{\frac{1-\bar{m}}{2}}F\right)$ is the new state of the Pauli frame. Since the state of the Pauli frame $F$ is not resolved until the correction qubit measurement outcome $m_4$ has been decoded, in the anti-commutation case, the basis of the subsequent correction qubit cannot be resolved earlier than $\gamma_\text{QM}$ after the measurement of the previous correction qubit, as illustrated in \cref{fig:pi8_diagram}.

\section{Noise Model}
\label{sec:noise-model}
 For our simulations, we employ the \emph{physical depolarizing} noise model, which captures the differing error rates of each operation in a quantum processor~\cite{Mohseni2026How, 1qbit2024topqad}. In a noiseless circuit, each ideal operation is either prepended or postpended by a noise channel, whose strength is determined by relevant theory that connects it to the infidelity of the operation. The noise channels are described in \cref{tab:noisechannels,tab:noisemodel}. The error rate of each operation can be determined by experimental benchmarking. However, our simulations employ the \emph{target} hardware parameters provided in \cref{tab:target-params}, a set of parameter values that are believed to be achievable in superconducting systems in the near term.
 
\begin{table}[h]
    \centering
    \begin{tabular}{ccc}
    \toprule
    Name & Operation & Constraint\\
    \midrule
    One-qubit depolarizing $\mathcal{D}_1(p)$ & $(1-p)\rho + \frac{p}{3}(X\rho X + Y\rho Y + Z\rho Z)$ & $0 \le p \le 0.5$ \\[0.5em]
    Two-qubit depolarizing $\mathcal{D}_2(p)$ & $(1-p)\rho + \frac{p}{15}\sum_{P \in \{I,X,Y,Z\}^{\otimes 2}\setminus II} P\rho P$ & $0 \le p \le 0.75$ \\[0.5em]
    One-qubit bit-flip channel $\mathcal{F}_X(p)$ & $(1-p)\rho + pX\rho X$ & $0 \le p \le 1$ \\ 
    \bottomrule
    \end{tabular}
    \caption{The set of noise channels used to describe the noise model in \cref{tab:noisemodel}. Each operation is a Pauli channel; hence, it can be simulated by a Clifford simulator.}
    \label{tab:noisechannels}
\end{table}

\begin{table}[h]
    \centering
    \begin{tabular}{cc}
    \toprule
    Operation $\mathcal{O}$ & Noisy operation \\
    \midrule
    One-qubit gate & $\mathcal{D}_1\left(\frac{3}{2}p_{g1}\right) \circ \mathcal{O}$ \\[0.5em]
    Two-qubit gate & $\mathcal{D}_2\left(\frac{5}{4}p_{g2}\right) \circ \mathcal{O}$ \\[0.5em]
    State preparation in $\ket{0}$ & $\mathcal{F}_X(p_p) \circ \mathcal{O}$ \\[0.3em]
    Reset to $\ket{0}$ & $\mathcal{F}_X(p_r) \circ \mathcal{O}$ \\[0.5em]
    Measurement in $Z$ basis& $\mathcal{O} \circ \mathcal{F}_X(p_m)$ \\[0.5em]
    Idle on qubit for time $t$ & $\mathcal{D}_1\left(\frac{3}{4}(1 - e^{-t / T_1})\right)$ \\
    \bottomrule
    \end{tabular}
    \caption{The physical depolarizing noise model~\cite{Mohseni2026How, 1qbit2024topqad}. Each operation in the noiseless circuit is replaced by the corresponding noisy operation. The notation $\mathcal{B} \circ \mathcal{A}$ means $\mathcal{B}$ is executed after $\mathcal{A}$. The noise channels used are specified in \cref{tab:noisechannels}. An example set of numerical values that can be used to calculate the strengths of these noise channels are specified in \cref{tab:target-params}. Other sets are available in Ref.~\cite{Mohseni2026How}.}
    \label{tab:noisemodel}
\end{table}

\begin{table}[h]
\begin{center}
    \begin{tabular}{cc}
    \toprule
    Hardware parameter & Value \\
    \midrule
    One-qubit gate error $p_{g1}$ & 0.0002 \\
    Two-qubit gate error $p_{g2}$ & 0.0005 \\
    State preparation error $p_p$ & 0.01 \\
    Reset error $p_r$ & 0.005 \\
    Measurement error $p_m$ & 0.005 \\
    $T_1$, $T_2$ times & \SI{200}{\micro\second} \\
    One-qubit gate time $t$ & \SI{25}{\nano\second} \\
    Two-qubit gate time $t$ & \SI{25}{\nano\second} \\
    State preparation time $t$ & \SI{1}{\micro\second} \\
    Reset time $t$ & \SI{100}{\nano\second} \\
    Measurement time $t$ & \SI{100}{\nano\second} \\
    \bottomrule
    \end{tabular}
    \end{center}
    \caption{Set of gate errors and times, and qubit relaxation and coherence times used in our simulations. These hardware parameters are believed to be achievable in the near term for superconducting quantum processors, and are hence referred to as \emph{target} hardware parameters~\cite{Mohseni2026How}. They serve as input to the physical depolarizing noise model specified in \cref{tab:noisemodel}.}
    \label{tab:target-params}
\end{table}

\section{FTQC Protocol Performance}
\label{sec:ftqc-performance}

Using the noise model defined in Appendix~\ref{sec:noise-model}, we characterize the four fundamental logical operations used in the LMPSC logical microarchitecture: quantum memory, growth, magic state preparation, and lattice surgery~\cite{Litinski2019Game,Fowler2019Low}. Each operation acts on one or more patches of size $d \times d$ and takes exactly $d$ stabilization rounds, and hence a logical cycle time of $\tau_\text{logical} = d~\unit{\micro\second}$. This ensures synchronization of all operations in a single module of an LMPSC microarchitecture. 

As reported previously~\cite{Mohseni2026How}, under the chosen noise model, the logical error rate of quantum memory is given by \cref{eq:ler_memory}, with $\mu = 0.019(4)$ and $\Lambda = 9.3(3)$. Due to the similarity in the protocols, we approximate the error rate of the growth protocol by the quantum memory protocol. We select magic state cultivation as the magic state preparation protocol due to its impressive performance~\cite{Gidney2024Magic}. This is a non-deterministic protocol, whose logical error rate for distances $d \le 25$ is $\text{Pr}_\text{magic} \le 4.73 \times 10^{-5}$ and discard rate is $0.41$~\cite{Mohseni2026How}.

We estimate the performance of lattice surgery by numerically calculating the fitting parameters $\mu_S, \Lambda_S, \mu_T$, and $\Lambda_T$ found in \cref{eq:ler_ls_simplified}. Using Stim~\cite{Gidney2021Stim} and PyMatching~\cite{Higgott2025Sparse}, we perform simulations in which two logical qubits, labelled $0$ and $1$, encoded in two surface code patches of size $d \times d$, separated by one bus patch, also of size $d \times d$, undergo a $Q=X \otimes X$ lattice surgery. Two simulations are required to estimate the four fitting parameters.

\begin{figure*}[t!]
    \centering
    \begin{subfigure}[b]{0.48\textwidth}
        \centering
        \includegraphics[width=\textwidth]{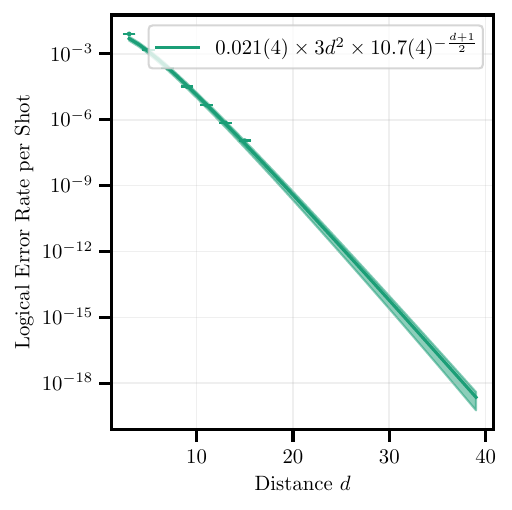}
        \caption{Distance scaling}
        \label{fig:ls-distance-scaling}
    \end{subfigure}
    \hfill
    \begin{subfigure}[b]{0.48\textwidth}
        \centering
        \includegraphics[width=\textwidth]{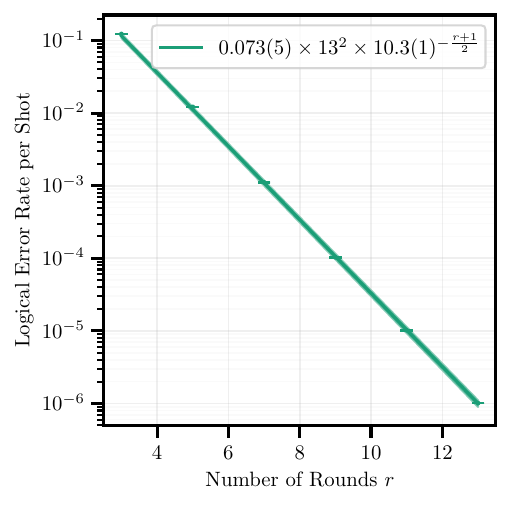}
        \caption{Round scaling}
        \label{fig:ls-rounds-scaling}
    \end{subfigure}
    \caption{Logical error rates of lattice surgery (a) as a function of the distance $d$, and (b) as a function of the number of rounds $r$. Data points are shown with error bars that have a range too small to be discernible. The data is fit to \cref{eq:ler_ls_distance} and \cref{eq:ler_ls_rounds}, respectively. The uncertainties in the fits, depicted by the shaded regions, is small at the distances used in the optimized microarchitectures.}
    \label{fig:lattice-surgery-results}
\end{figure*}

In the first simulation, we determine $\mu_S$ and $\Lambda_S$ by setting $r=d$ and varying $d$. First, the two logical qubit patches are prepared in $\ket{00}_L$, and the bus data qubits are prepared in $\ket{0}$. Next, a merged code is created by performing $r$ merged stabilization rounds on the three patches. Then, the two codes are split by measuring the bus in the $Z$ basis. The expected output state is
\begin{equation}
\frac{(I+mXX)}{\sqrt{2}}\ket{00}_L = \frac{\ket{00}_L + m\ket{11}_L}{\sqrt{2}},
\end{equation}
where $m$ is the lattice surgery outcome. To validate this state, we measure all data qubits in the $Z$ basis and determine the logical $Z$ operator for the merged code, $Z_{L,0} \otimes Z_{L,1} \otimes Z_{L,\text{bus}}$. This operator is susceptible to logical $X$ errors, but does not detect time-like failures. Hence, the leading-order term in 
\cref{eq:ler_ls_simplified} is
\begin{equation}
    \text{Pr}_\text{LS}(d, d, 2, 1) \approx \mu_S \times 3d^2 \times \Lambda_S^{-\frac{d+1}{2}}. \label{eq:ler_ls_distance}
\end{equation}
In the second simulation, we determine $\mu_T$ and $\Lambda_T$ by setting $d=13$ and varying $r$. The logical qubit patches are prepared in $\ket{+0}_L$, which leads to the expected output state of
\begin{equation}
    \ket{+0}_L + m \ket{+1}_L = \begin{cases} \ket{++}_L \text{ if } m=+1, \\ \ket{+-}_L \text{ if } m=-1.\end{cases}
\end{equation}
To validate this state, the data qubits of the second computation qubit are measured in the $X$ basis, from which the value of the operator $X_{L,1}$ can be inferred. While the value of this operator is random, its sum with $m$ should be $0$, unless a time-like error has flipped the value of $m$. Note that $X_{L,1}$ is sensitive to logical $Z$ errors, which have negligible probability in this simulation. Hence, the logical error rate is
\begin{equation}
    \text{Pr}_\text{LS}(13, r, 2, 1) \approx \mu_T \times 13^2 \times \Lambda_T^{-\frac{r+1}{2}}. \label{eq:ler_ls_rounds}
\end{equation}
The results of these simulations are shown in \cref{fig:lattice-surgery-results}. The values of $\Lambda_S$ and $\Lambda_T$ are close because $r=d$ is close to the optimal number of rounds required for minimal lattice surgery errors~\cite{Mohseni2026How}.

\end{appendices}

\bibliography{main}

\end{document}